\documentclass[preprint,12pt]{elsarticle} 
\biboptions{sort&compress}
\usepackage{amssymb,amsbsy,amsthm,amsmath,amsfonts,amssymb,amscd}
\usepackage{geometry}
\usepackage{bm} 
\usepackage{graphicx}
\usepackage{float}
\usepackage{multirow}
\usepackage{subfigure}
\usepackage{lineno}
\usepackage{hyperref}
\hypersetup{colorlinks, citecolor=black, filecolor=black, linkcolor=black, urlcolor=black}
\usepackage[ruled,linesnumbered]{algorithm2e}
\usepackage{appendix}
\usepackage{algpseudocode}
\usepackage{mathtools,nccmath}
\usepackage{adjustbox}
\usepackage{cleveref}
\graphicspath{{fig/}}
\usepackage{color}
\usepackage{soul}

\definecolor{aqua}{rgb}{0.0, 1.0, 1.0}

\definecolor{aquamarine}{rgb}{0.5, 1.0, 0.83}

\begin{document}

\begin{frontmatter}

\title{Damage identification for bridges using machine learning: Development and application to KW51 bridge}

\author[inst1,inst2]{Yuqing Qiu\corref{cor1}}\ead{yq2483@nyu.edu}
\author[inst2]{Bilal Ahmed}
\author[inst2,inst3]{Diab W. Abueidda\corref{cor1}}\ead{da3205@nyu.edu}
\author[inst2,inst4]{Waleed El-Sekelly}
\author[inst2]{Borja García de Soto}
\author[inst2]{Tarek Abdoun}
\author[inst1]{Hongli Ji}
\author[inst1]{Jinhao Qiu}
\author[inst2]{Mostafa E. Mobasher\corref{cor1}}\ead{mostafa.mobasher@nyu.edu}

\affiliation[inst1]{
    organization={The State Key Laboratory of Mechanics and Control of Mechanical Structures},
    addressline={Nanjing University of Aeronautics and Astronautics}, 
    country={China}
    }

\affiliation[inst2]{
    organization={Civil and Urban Engineering Department},
    addressline={New York University Abu Dhabi}, 
    country={United Arab Emirates}
}
    
\affiliation[inst3]{
    organization={National Center for Supercomputing Applications},
    addressline={University of Illinois at Urbana-Champaign}, 
    country={United States of America}
}

\affiliation[inst4]{
    organization={Department of Structural Engineering},
    addressline={Mansoura University}, 
    country={Mansoura, Egypt}
}

\cortext[cor1]{Corresponding author}

\begin{highlights}
\item A comprehensive machine learning based damage identification (CMLDI) method is developed.
\item A novel combination of signal processing and machine learning methods is proposed.
\item The resulting approach successfully captures the existence, magnitude, and location of the damage.
\item The CMLDI framework can work with short-term and long-term time history data.
\item The proposed CMLDI method is applied to a real structure, the KW51 bridge.

\end{highlights}

\begin{abstract}
The available tools for damage identification in civil engineering structures are known to be computationally expensive and data-demanding. This paper proposes a comprehensive machine learning based damage identification (CMLDI) method that integrates modal analysis and dynamic analysis strategies. The proposed approach is applied to a real structure - KW51 railway bridge in Leuven. CMLDI diligently combines signal processing, machine learning (ML), and structural analysis techniques to achieve a fast damage identification solver that relies on minimal monitoring data. CMLDI considers modal analysis inputs and extracted features from acceleration responses to inform the damage identification based on the long-term and short-term monitoring data. Results of operational modal analysis, through the analysis of long-term monitoring data, are analyzed using pre-trained k-nearest neighbor (kNN) classifiers to identify damage existence, location, and magnitude. A well-crafted assembly of signal processing and ML methods is used to analyze acceleration time histories. Stacked gated recurrent unit (Stacked GRU) networks are used to identify damage existence, kNN classifiers are used to identify damage magnitude, and convolutions neural networks (CNN) are used to identify damage location. The damage identification results for the KW51 bridge demonstrate this approach's high accuracy, efficiency, and robustness. In this work, the training data is retrieved from the sensor of the KW51 bridge as well as the numerical finite element model (FEM). The proposed approach presents a systematic path to the generation of training data using a validated FEM. The data generation relies on modeling combinations of damage locations and magnitudes along the bridge.
\end{abstract}

\end{frontmatter}

\section{Introduction} \label{Section1-Introduction}
Damage identification methods for bridge structural health monitoring (SHM) have garnered significant attention over the years \cite{Development-bridges}. This increased interest is primarily driven by the aging infrastructure worldwide. More frequent inspections are necessary to ensure safety as these structures age and deteriorate. This challenge is magnified by the sheer number of bridges: Europe alone has approximately one million highway bridges, 35\% of which are over 100 years old \cite{Bridges-Europe}. Meanwhile, the situation is similar in the United States \cite{Bridges-US-2021} and China \cite{Bridges-China-2024}. Therefore, research on damage identification methods for bridge structures is crucial \cite{bridge-Portugal, Tsing-Ma-bridge, bridge-Greece}. According to Rytter \cite{Doctor_damage_detection}, SHM fundamentally involves detecting damage presence,  location, and magnitude and predicting structural life. This paper leverages machine learning and artificial intelligence advancements to develop a systematic damage identification approach for full-scale bridge structures. 

\subsection{Existing bridge damage identification approaches}\label{Existing bridge damage identification approaches}

Numerous classical SHM approaches for bridge structures have been proposed \cite{ SHM-review-2016}, examining the effects of various loads (static load \cite{Vibration-based-SHM}, dynamic load \cite{SHM-Displacement-Stiffness}, wind load \cite{SHM-FullScale}, and temperature load \cite{SHM-Application}) on the reliability of bridge structures. For instance, Catbas et al. \cite{SHM-Application} conducted a reliability assessment for a long-span truss bridge, while Betti et al. \cite{Betti-Bridge,betti1999conditions,Wind-Analysis} investigated damage detection in cable-suspended bridges. These approaches analyzed measured data to extract structural parameters and assess the structural characteristics of both the bridge structural components \cite{SHM-Stiffness-Vibration,SHM-MEMS, Bridge-Cables}. Most traditional SHM approaches rely on extensive datasets from long-term monitoring \cite{SHM-Civil,  AI-SHM-bridges}. However, these approaches are typically data-driven \cite{SHM-building-bridge, SHM-Damage-limited}; hence, they require the availability of significant monitoring datasets \cite{Review-of-Modal-Based-Damage-Detection}. Moreover, traditional analysis methods typically rely on signal processing techniques, which may not always be physics-informed, making extracting substantial information on structural damage challenging, especially for full-scale civil engineering applications \cite{Review-case-study}. 

To enhance the reliability of SHM methods, finite element models (FEM) are increasingly used, particularly when monitoring data is limited. Incorporating FEM into SHM \cite{FEM-HighFidelity} introduces the underlying physics and knowledge of the structure's stiffness and mass, providing a foundation for advanced damage identification methods. Moreover, FEM improves the overall effectiveness and efficiency of SHM systems \cite{SHM-FEM-DL}. Examples of FEM-based SHM include using FEM to simulate the structural damage \cite{SHM-Bayesian} and FEM updating method for monitoring railway bridges \cite{SHM-Displacement-Stiffness}. The FEM-based SHM methods are physics-informed, allowing them to rely on limited monitoring data \cite{SHM-Accelerator}. Validated FEM models can also be used to simulate various structural damage scenarios and form a prior understanding of possible structural failure mechanisms \cite{SHM-FEM-Experiment}. On the other hand, validated and detailed FEM models are typically complex, difficult to develop and validate, and computationally expensive, particularly when non-linear failure modeling is sought \cite{FEM-HighFidelity, SHM-FEM-DL,SHM-Accelerator}. 

Several researchers have attempted to use machine learning (ML) for damage identification applications to address the aforementioned challenges \cite{review-ML-damage}. ML algorithms, such as those presented in  \cite{2D-CNN-time,Hilbert, MEMD,wavelet-AE-CNN,simpleCNN}, are remarkably powerful in the solution of inverse problems. ML method can effectively model physical phenomena using only sampled data from measurements or simulations; moreover, they have the major advantage of extremely low inference time \cite{SHM-review-ML}. This capability is particularly useful when the underlying phenomenon is unknown or simulations are computationally expensive. 

Generally, data-driven methods for damage detection incorporate two major data sources: 1 - natural frequencies \cite{frequency-mssp-2005,Frequency-KNN-mssp, JSP-frequency-CNN}, and 2 - time series \cite{LSTM-CNN-time, DI-GRU,SHM-GAN, SHM-ANN}. The natural frequency is an inherent characteristic of the structure and is not affected by dynamic loads, so it is an ideal characteristic value for structural damage detection \cite{ANN-damageDetection-FrequencyFilter}. The measurement of natural frequency is time-consuming and requires long-term monitoring data. Therefore, it is also necessary to use time series damage detection \cite{time-series-3D-CNN}. Many researchers have focused on ML algorithms using dynamic time series input , including long short term memory (LSTM) networks \cite{LSTM-CNN-time}, gated recurrent unit (GRU) networks \cite{DI-GRU}, and artificial neural network (ANN) \cite{SHM-ANN}, and so on. Additionally, the wavelet transform \cite{wavelet-packet-energy} showed significant efficiency for transforming the time series into time-frequency domain images which can then be analyzed using convolution neural network (CNN) \cite{simpleCNN, 2D-CNN-time}. The above-mentioned methods face two major limitations. The first is the training data availability, this limitation is typically resolved either by using numerical models to generate fictitious data \cite{SHM-Big-Data,SHM-AI, Soft-computing, FEM-SHM-review}. The second is the limited application and applicability of these methods to simplified structures, which makes the extension to real-world structures a persistent challenge. 

Some of the ML approaches to system and damage identification have been applied to real-world structures. Examples include the applications to bridges \cite{Hybrid-DI, DL-bridge-FEM} and concrete buildings \cite{DL-FEM-updating, CNN-Concrete-Building, Damage-location-quantify}. Particularly, these recent advances have made solid steps towards staged and methodical damage identification that predicts damage existence, location, and magnitude. However, the results in these studies showed limited accuracy, which limits the reliability of these methods. Moreover, these methods typically rely on one type of data input, either short-term time series or long-term based modal analysis results. They also make minimal utilization of the existing body of knowledge in signal processing and structural health monitoring techniques. Based on the above, it can be concluded that there is a need for an efficient and reliable damage identification methodology that can utilize multiple types of data-input and leverage the existing ML and signal processing methods. There is also a need for well-qualified methods that can be efficiently applicable to full-scale real world structures. 

\subsection{Problem statement and overall approach}\label{Problem statement and overall approach}
\begin{figure}[!htb]
    \centering
    \includegraphics[width=1\textwidth]{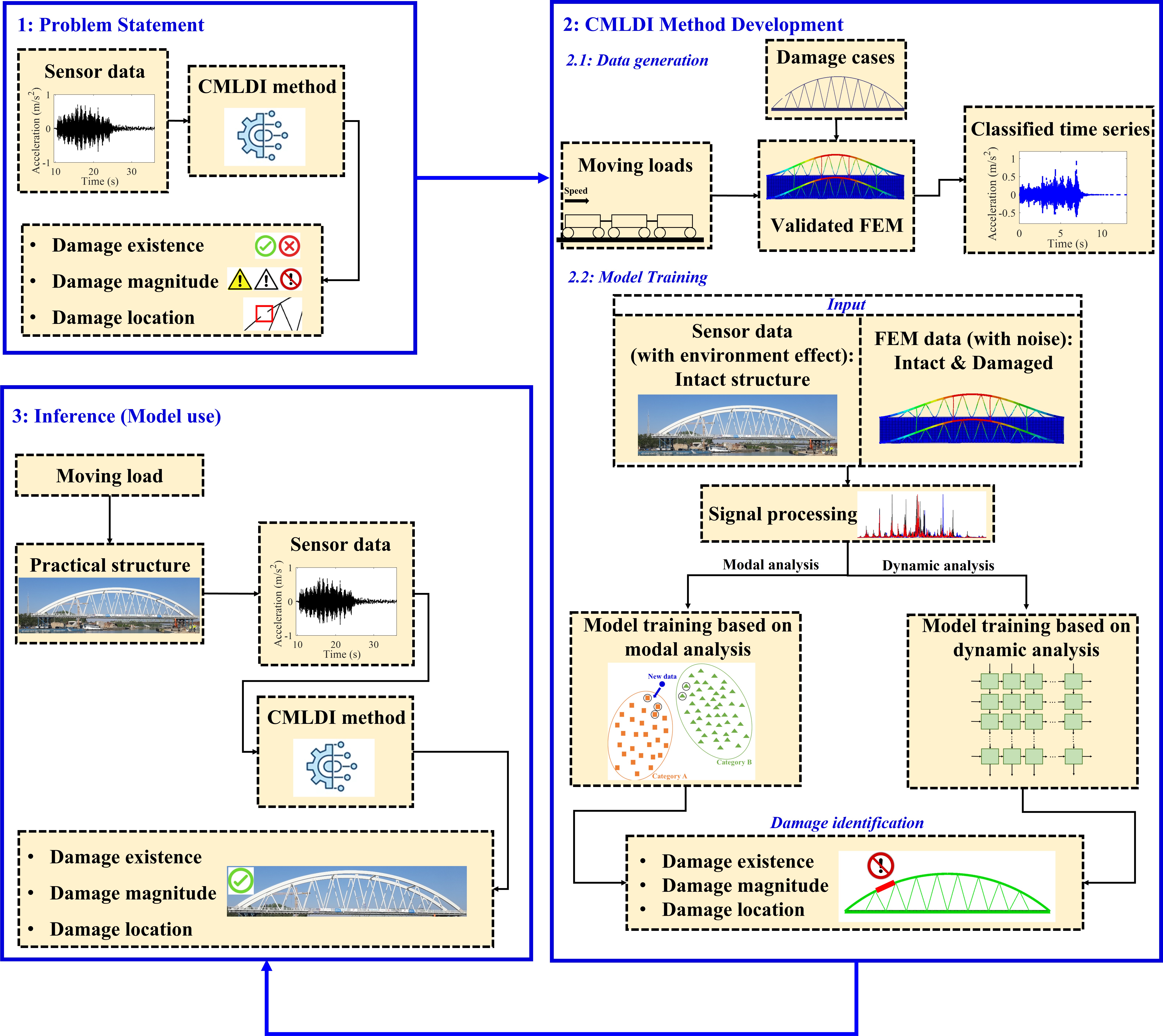}
    \caption{The overall approach including problem statement, CMLDI method development and inference}
    \label{Overall}
\end{figure}

This paper proposes a comprehensive machine learning-based damage identification (CMLDI) method. This approach attempts to introduce comprehensive damage identification in bridges, with the aim of having a real-time accurate analysis that can be directly used by engineers and asset owners. The problem statement and overall approach are illustrated in Figure \ref{Overall}. In this paper, we develop and apply the CMLDI approach to the KW51 railway bridge in Belgium, which has been investigated and modeled in several previous studies \cite{KW51-NaturalFrequency-MSSP, KW51-MSSP-stresses, KW51-classify-Before-After, KW51-GraphSignal2024}. The steps involved in the overall approach are outlined as follows:
\begin{enumerate}
\item \textbf{Problem Statement:}
    \begin{itemize}
    \item There is a need to develop a fast and accurate damage identification model for practical full-scale structures.
    \item The model should take the SHM data and input and output the damage existence, location, and magnitude.
    \end{itemize}
\item \textbf{CMLDI Method Development:}
    \begin{itemize}
    \item \textbf{Data collection and generation:} We utilize the available monitoring data to represent the intact structure. We develop a FEM model and validate it against the monitoring data. The validated FEM is then utilized to simulate structural damage cases and generate virtual time histories representing these cases. We simulate multiple damage scenarios, including various damage magnitudes and locations.
    \item \textbf{CMLDI model training: } The CMLDI model is trained for both long-term and short-term monitoring using on two strategies: modal analysis and dynamic analysis. Input data include both sensor data and FEM data.
    \end{itemize}
\item \textbf{Inference (Model use): }
    \begin{itemize}
    \item Once trained, the CMLDI model can be effectively used for the damage identification of the bridge structure.
    \item The input to the CMLDI model is practical sensor data, and the outputs are the existence, magnitude, and location of structural damage.
    \end{itemize}    
\end{enumerate}

Compared to the existing literature, this paper presents several novel contributions:
\begin{itemize}
    \item A comprehensive damage identification model that can output the triad of damage existence, location, and magnitude based on SHM data.
    \item A multi-staged approach that diligently combines various signal processing and machine learning methods to identify damage aspects. This approach can be directly extended to other civil and mechanical engineering applications.
    \item An approach to utilize short-term and long-term monitoring data to infer the damage status of the structure.
    \item A methodical approach to designing damage scenarios that can be used as a source for training data based on FEM simulations.
    \item The application of the proposed approach to real full-scale bridge monitoring while utilizing available monitoring data. The results showcase the efficiency and accuracy of the proposed method. 
\end{itemize}   

The layout of this paper is as follows: Section \ref{Section2-Methodology} explains the proposed CMLDI method and provides an overall flowchart. Section \ref{Section3-FEM} presents the details of FEM and the approaches to generating data. Sections \ref{Section4-Signal processing} and \ref{Section5-ML} covers the signal processing approaches and the machine learning methods, respectively. Section \ref{Section6-frequency} presents the damage identification results based on modal analysis. Section \ref{Section7-acceleration} applies signal processing techniques for damage identification, considering dynamic analysis.Finally, Section \ref{Section8-Discussion} presents the discussion, conclusions, and future directions.

\section{Methodology}\label{Section2-Methodology}

\begin{figure}[!htb]
    \centering
    \includegraphics[width=1\textwidth]{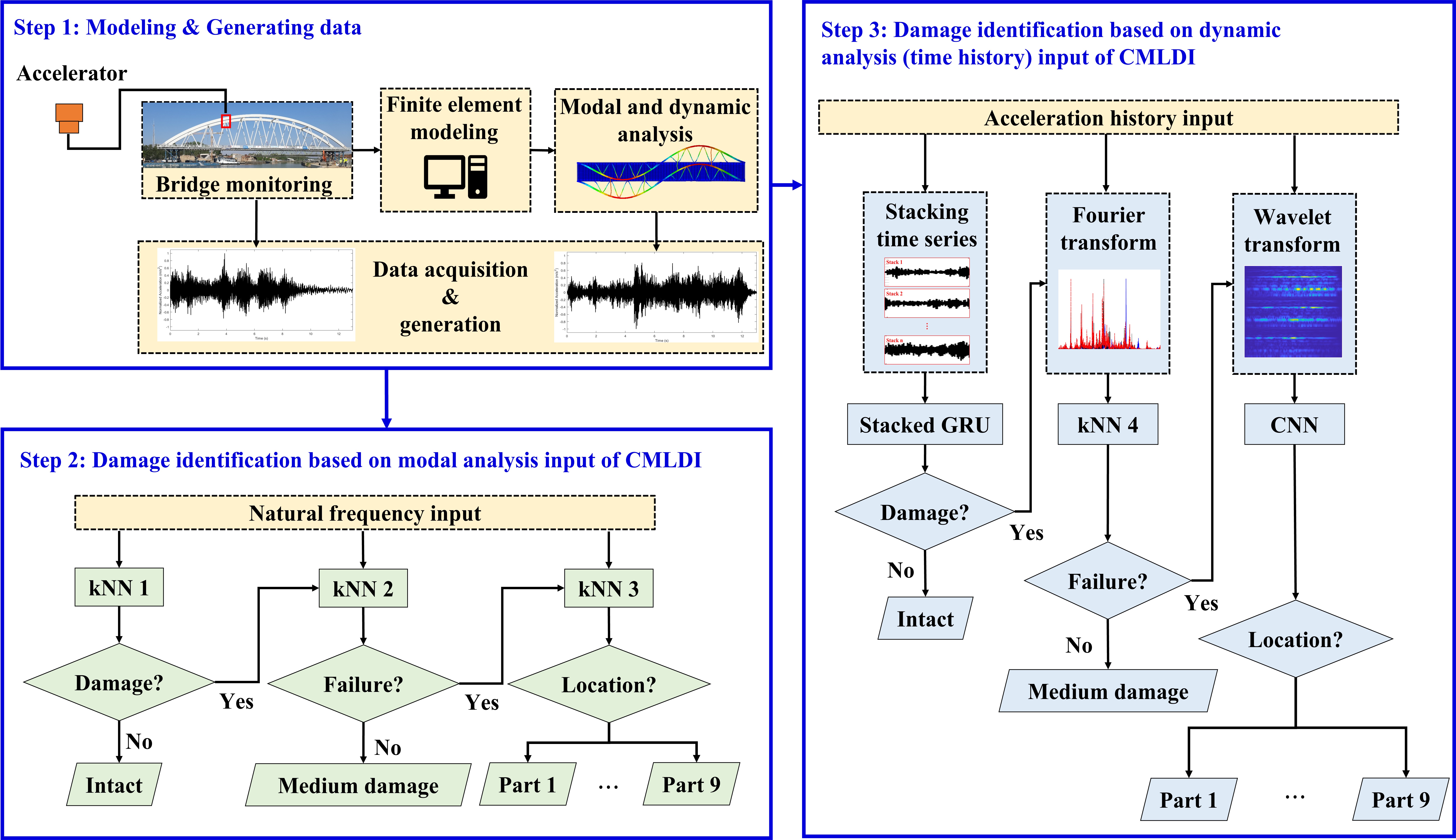}
    \caption{The flowchart of CMLDI method for the bridge structure}
    \label{FlowChartExplain}
\end{figure}

This paper proposes the CMLDI method, which employs a limited number of sensors for both long-term and short-term monitoring, to meet the comprehensive requirements of damage detection in civil engineering. The flowchart of CMLDI method is shown in Figure \ref{FlowChartExplain}. The method is broadly divided into three main steps:

\begin{enumerate}
\item \textbf{Data generation}: 
\begin{itemize}
\item \textbf{FEM development:} The FEM model is developed to simulate the structural response. 
\item \textbf{Data generation:} Several design scenarios are designed, implemented, and then used to generate data representing different damage magnitudes and locations. 
\end{itemize}

\item \textbf{Damage identification based on modal analysis input:} Three k-Nearest Neighbor (kNN) classifiers are developed to identify damage existence, location, and magnitude.
\item \textbf{Damage identification based on acceleration (time history) input:} 

\begin{itemize}
\item \textbf{Damage existence identification:} Extremely long temporal sequence is stacked by stacking approach. The stacked time series is fed into the stacked GRU network to determine damage existence. 
\item \textbf{Damage magnitude identification:} Fourier transform is applied to get the frequency features of accelerations. The frequency features are used to identify the severity of damage by the kNN-4 classifier.
\item \textbf{Damage location identification:} Wavelet transform is proposed to obtain the time-frequency image. A CNN is designed to classify these images to obtain the failure location.
\end{itemize}

\end{enumerate}

It is worth noting that the three machine learning algorithms are specified in the order in which they exhibit the most powerful performance. The order in which the three machine learning methods are specified depends on the type of input and the efficiency.

In the proposed approach, it is assumed that the bridge damage is the main variant leading to significant changes in the bridge natural frequencies and acceleration signals. Environmental effects are partially accounted for by using available sensing data and introducing artificial noise to the FEM-generated data. However, a more substantial consideration of environmental effects would require long-term monitoring data that can capture changes in factors like temperature and humidity over several months to years, which is currently not feasible. Variations of the load are not accounted for in this study.

\begin{figure}[!htb]
    \centering
    \includegraphics[width=0.35\textwidth]{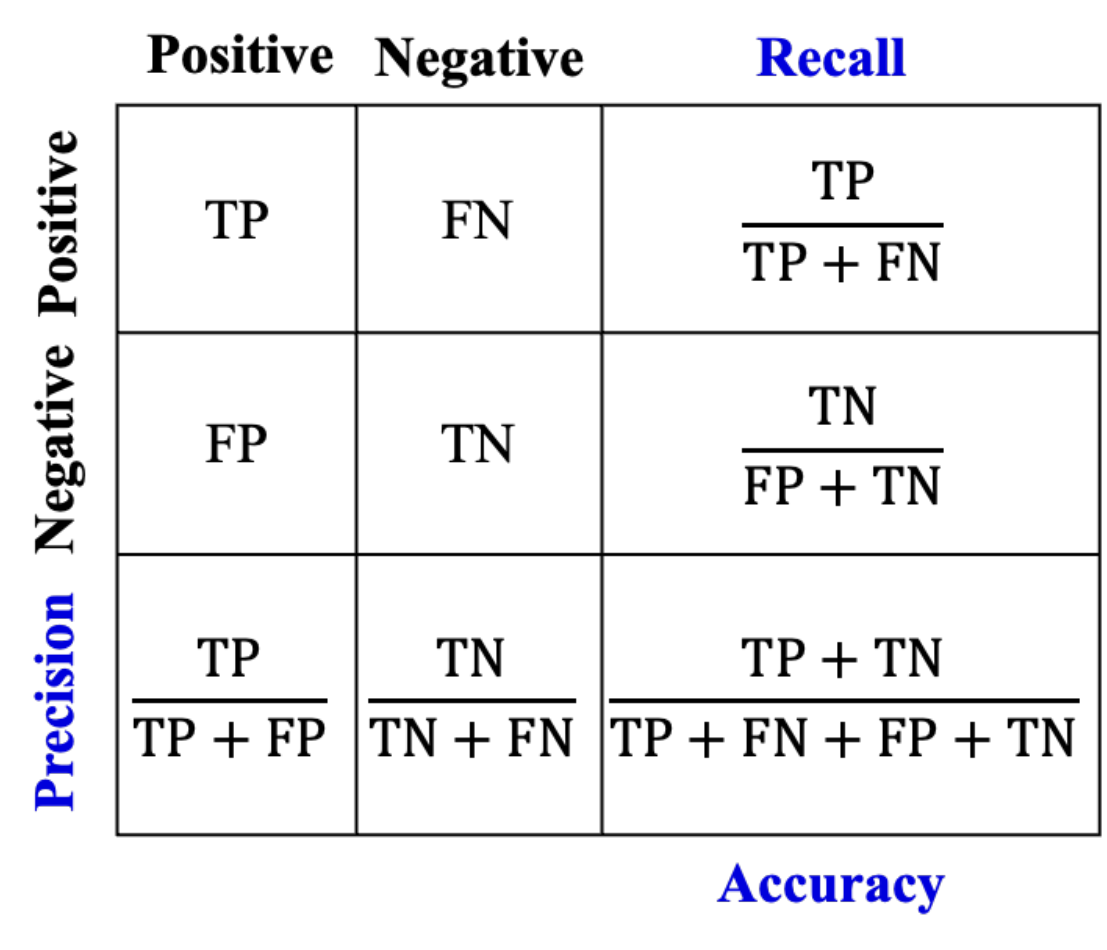}
    \caption{The schematic of the confusion matrix}
    \label{ConfusionMatrix}
\end{figure}

The proposed method is applied to the KW51 bridge. The ML approaches employed within the CMLDI method are classifiers, and the confusion matrix \cite{Generalized-stacked-LSTM} is adopted to display the accuracy of predicted classes. A schematic of the confusion matrix is shown in Figure \ref{ConfusionMatrix}. The diagonal terms of the confusion matrix represent correctly predicted values, while the off-diagonal terms represent the incorrectly predicted values. The overall accuracy of the network is indicated as a percentage in the matrix entry at the intersection of the bottom row and the rightmost column. The performance of a classifier can be assessed using the bottom row for precision and the rightmost column for recall. Precision measures the proportion of predicted positive observations that are actually positive that is in the bottom row of the confusion matrix. Recall measures the proportion of actual positive observations that are correctly predicted, which is found in the rightmost column of the confusion matrix. Thus, the following inferences can be made:
\begin{enumerate}
    \item \textbf{Diagonal Terms:} High values indicate good classification performance.
    \item \textbf{Off-Diagonal Terms:} High values indicate areas where the model needs improvement.
    \item \textbf{Bottom Row (Precision):} High precision indicates that most of the predicted positives are actual positives.
    \item \textbf{Rightmost Column (Recall):} High recall indicates that the model is correctly identifying a large portion of the actual positives.
\end{enumerate}

\section{Finite element model and data generation}\label{Section3-FEM}

\begin{figure}[!htb]
    \centering
    \includegraphics[width=0.5\textwidth]{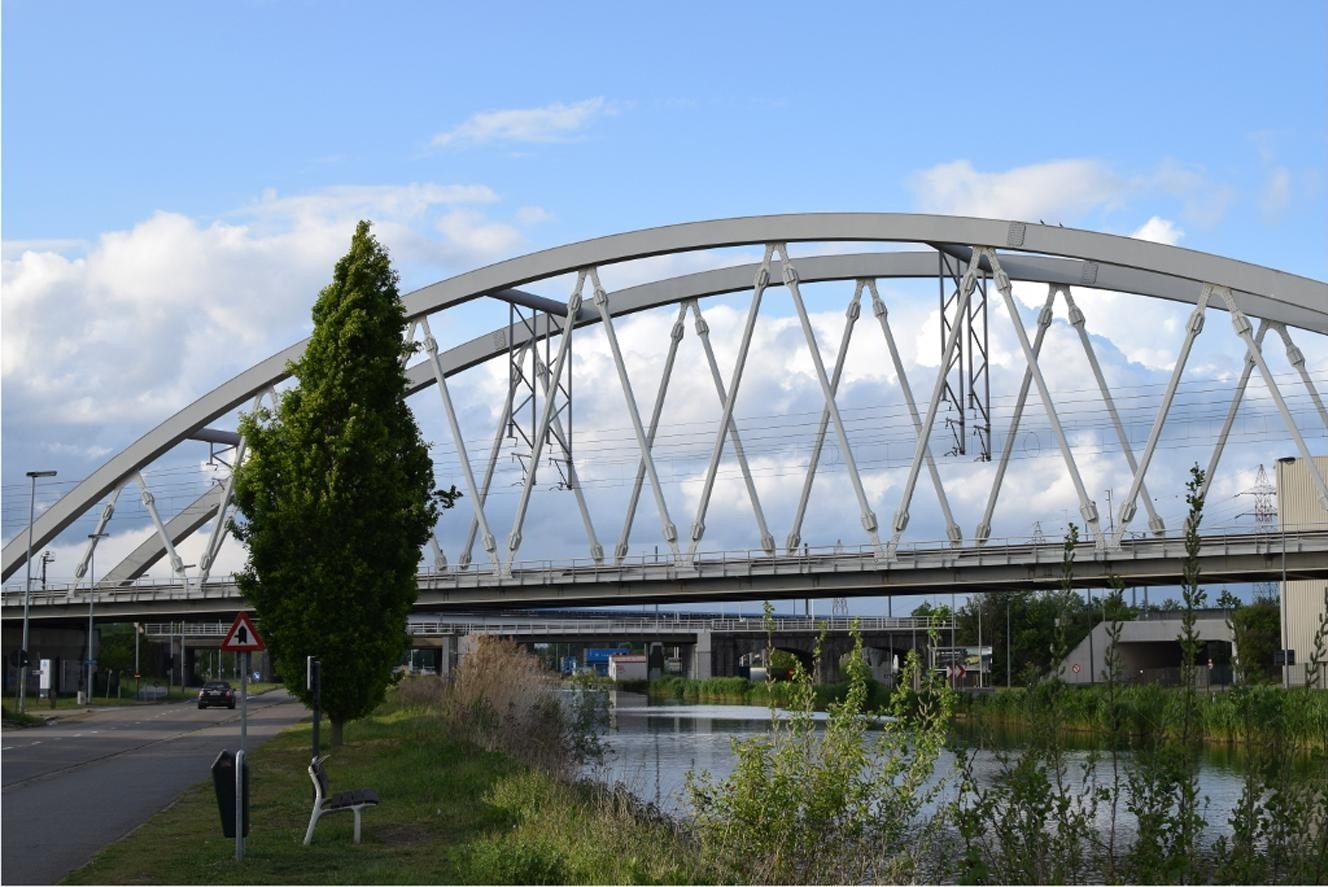}
    \caption{Railway bridge KW51 (\citep{Structurae}) in Leuven, Belgium}
    \label{KW51-bridge}
\end{figure}

KW51 is a railway bridge spanning the Leuven-Mechelen canal near Leuven, Belgium. The bridge is part of the 100km Line 36, running from Brussels to Liège. The KW51 bridge is a steel, single-span tied arch bridge with a two-track deck suspended from the arch by thirty-two inclined braces, as shown in Figure \ref{KW51-bridge}. The deck is supported by two main girders, stiffened by thirty-three transverse beams. The girders are composed of three sections connected with steel plates. This bowstring-type bridge has a length of 115m and a width of 12.4m. It accommodates two ballasted tracks, referred to as track A (on the north side) and track B (on the south side) \cite{KW51-validation-frequency}. The tracks are curved, with radii of of 1125m for track A and 1121m for track B. Through long-term monitoring, the first 14 natural frequencies of the bridge structure were obtained using automated operational modal analysis \cite{KW51-Monitoring-Frequency}. This paper considers the moving load on track B, assuming a speed between 15 m/s and 30 m/s (54km/hr to 108 km/hr).

\begin{figure}[!htb]
    \centering
    \includegraphics[width=0.8\textwidth]{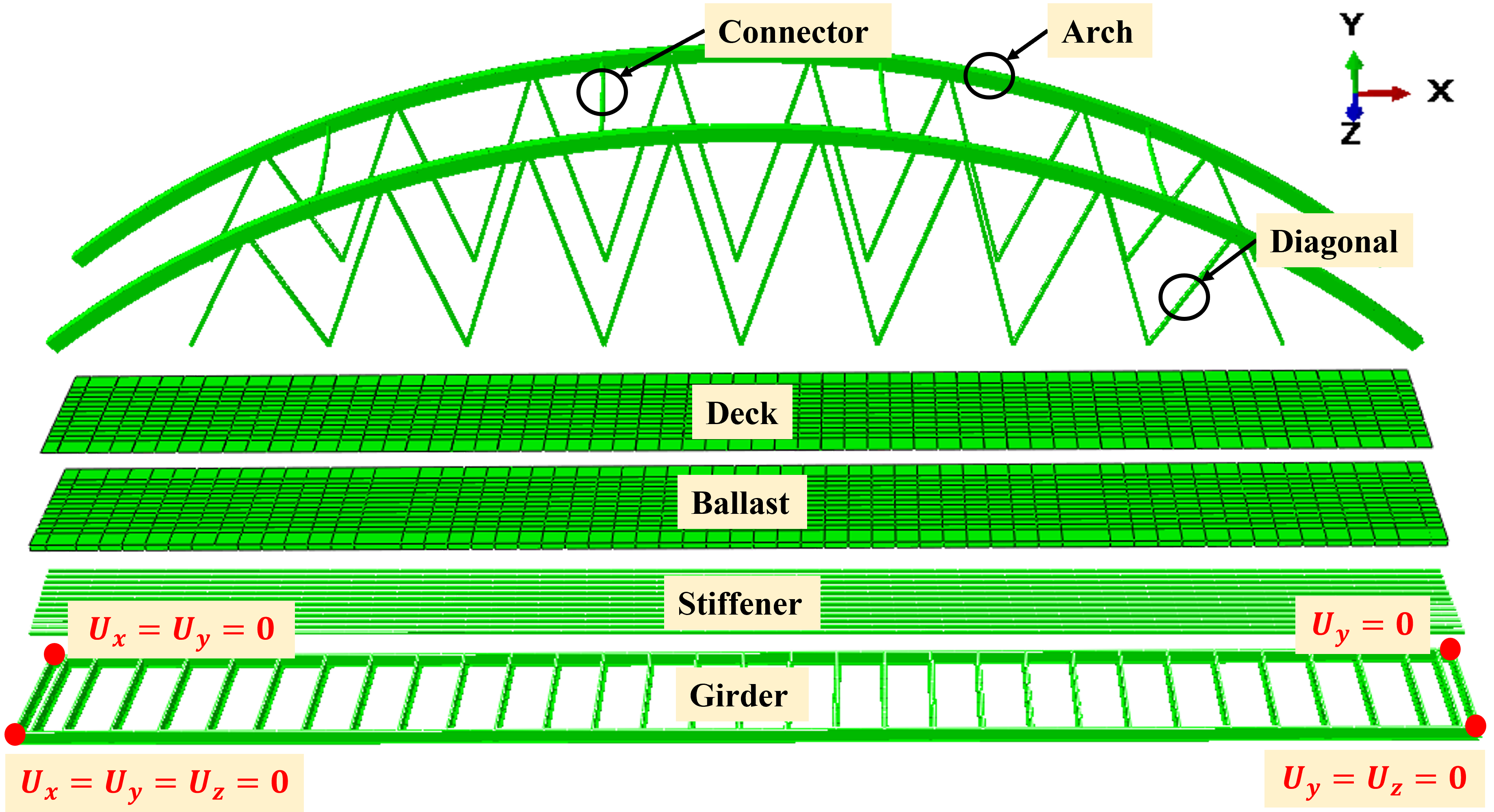}
    \caption{The components and boundary conditions of KW51 bridge}
    \label{KW51-instance}
\end{figure}

The Abaqus platform is used to establish the FEM of the KW51 bridge. The bridge structure consists of arches, connectors, diagonals, deck, ballast, U-shape stiffeners, and girders, as shown in Figure \ref{KW51-instance}. Tie constraints are used to connect the structural components. The boundary conditions are simulated by defining the four constraint locations of the girder, as illustrated in Figure \ref{KW51-instance}. The dimensional and material parameters for FEM of the KW51 bridge are listed in the Table \ref{Parameters of KW51}. Steel is used for all bridge components except the ballast.

\begin{table}[h]\footnotesize
\centering
\caption{The dimensional and material parameters for FEM of the KW51 bridge}
\label{Parameters of KW51}
\begin{tabular}{|c|c|}
\hline
\text{Description} & \text{Dimensions} \\ \hline
\text{Total length of deck}&\text{115m}\\
\text{Total width of deck} &\text{12.4m} \\
\text{Box arch section} &\text{Width=0.86m, Height=1.3m, Thickness=0.045m} \\
\text{Box diagonal section} &\text{Width=0.345m, Height=0.35m, Thickness=0.016m} \\
\text{Pipe connector section} &\text{Radius=0.2m, Thickness=0.002m} \\
\text{Deck section} &\text{Thickness=0.015m} \\
\text{Ballast section} &\text{Thickness=0.6m} \\
\text{U-shape stiffener section} &\text{Width=0.25m, Height=0.25m, Thickness=0.008m} \\
\text{T-shape girder section} &\text{Width=0.6m, Height=1.235m, Thickness=0.08m} \\
\hline
\text{Description} & \text{Materials} \\ \hline
\text{Steel}&\text{Density=7750kg/m$^{3}$, Young's modulus=210GPa}\\
\text{Steel damping}&\text{Alpha=0.7, Beta=0.5}\\
\text{Ballast}&\text{Density=1900kg/m $ ^{3} $ , Young's modulus=550MPa}\\
\text{Ballast damping}&\text{Alpha=0.7, Beta=0.5}\\
\hline
\end{tabular}
\end{table}

\subsection{Natural frequency validation}\label{Natural frequency validation}

\begin{figure}[!htb]
    \centering  \includegraphics[width=0.9\textwidth]{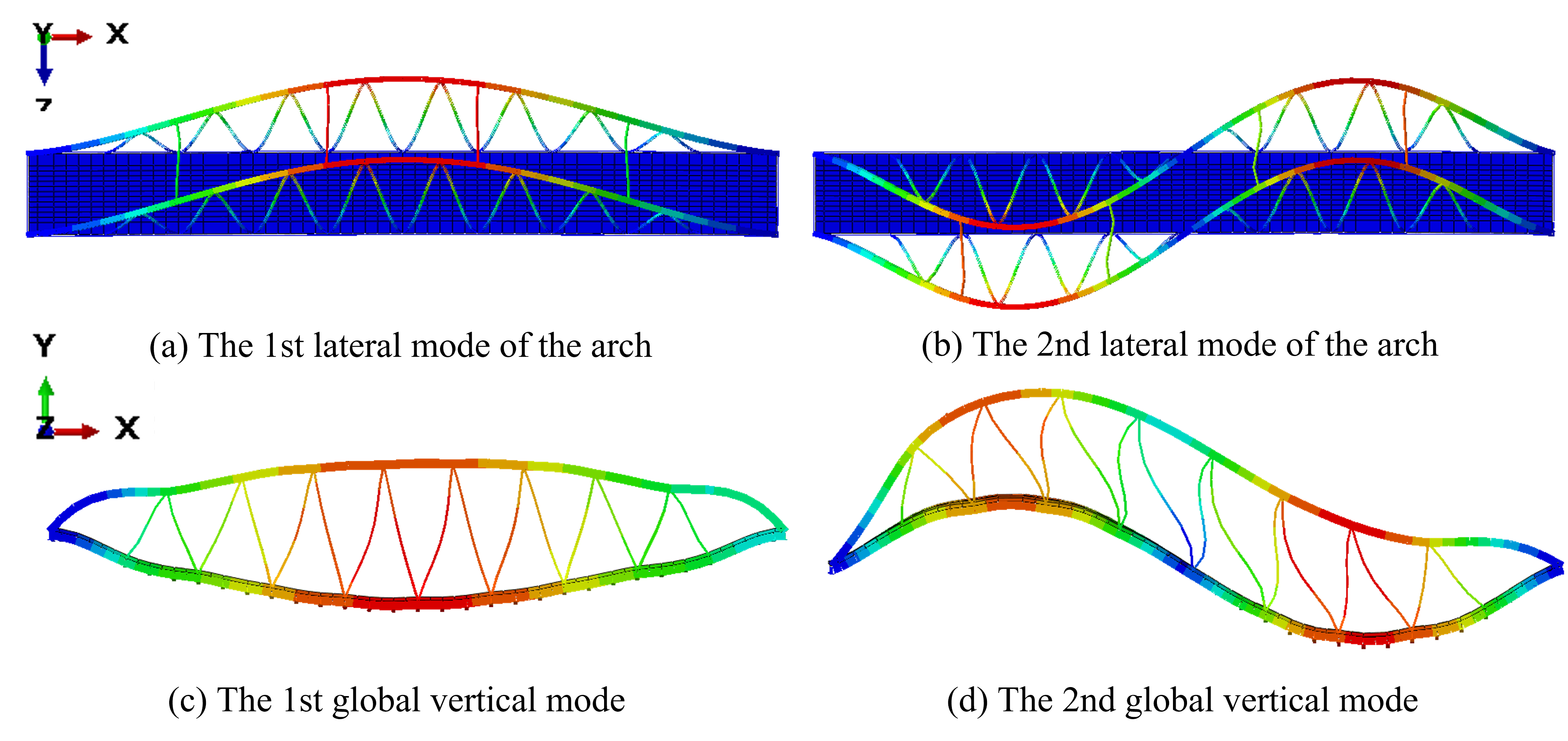}
    \caption{Mode shapes of FEM for KW51 bridge: (a) The 1st lateral mode of the arch; (b) The 2nd lateral mode of the arch; (c) The 1st global vertical mode; (d) The 2nd global vertical mode}
    \label{Mode shapes}
\end{figure}

The FEM is validated based on a set of natural frequencies determined through operational modal analysis (OMA) \cite{KW51-NaturalFrequency-MSSP}\cite{KW51-MSSP-stresses}. During the OMA, the longitudinal, transverse, and vertical modes were estimated by sensors on the bridge deck. Additionally, the lateral accelerations on the arch were measured at six locations. This comprehensive measurement captures detailed global modal shapes. Fourteen modes are retained, including five arch lateral bending modes and nine global modes involving both the arch and the deck motion. As an example, Figure \ref{Mode shapes} (a) and (b) show the first two arch lateral modes,  Figure \ref{Mode shapes} (c) and (d) show the global vertical modes obtained from the FEM. Table \ref{Comparison of natural frequency} compares the first 14 natural frequencies of the KW51 bridge obtained from FEM modal analysis and OMA experiments. Since measuring the natural frequencies of the bridge involves long-term monitoring, Table \ref{Comparison of natural frequency} presents the average values from 3000 sets of OMA measurements. The average accuracy of the FEM natural frequency is 93\%. 

\begin{table}[h]\footnotesize
\centering
\caption{Comparison of natural frequency between the FEM and measured values}
\label{Comparison of natural frequency}
\begin{tabular}{|c|c|c|c|}
\hline
\text{Description} & \text{FEM} & \text{Measured} & \text{Accuracy}\\ \hline
\text{1st lateral mode of the arches}&\text{0.55}&\text{0.51}& \text{91.24\%} \\
\text{2nd lateral mode of the arches} &\text{1.22} & \text{1.23}&\text{99.20\%} \\
\text{1st lateral mode of the bridge deck} &\text{1.73} & \text{1.87}&\text{92.50\%} \\
\text{1st global vertical mode} &\text{2.07} & \text{2.43}&\text{85.39\%} \\
\text{3rd lateral mode of the arches} &\text{2.02} & \text{2.53}&\text{79.53\%} \\
\text{2nd global vertical mode} &\text{2.78} & \text{2.92}&\text{95.22\%} \\
\text{4th lateral mode of the arches} &\text{3.21} & \text{3.55}&\text{90.49\%} \\
\text{1st global torsion} &\text{3.53} & \text{3.90}&\text{90.50\%} \\
\text{3rd global vertical mode} &\text{4.04} & \text{3.97}&\text{98.23\%} \\
\text{2nd global torsion} &\text{4.10} & \text{4.29}&\text{95.39\%} \\
\text{2nd lateral mode of the bridge deck and torsion mode} &\text{4.52} & \text{4.81}&\text{93.95\%} \\
\text{4th global vertical mode } &\text{5.28} & \text{5.31}&\text{99.35\%} \\
\text{3rd global torsion} &\text{6.11} & \text{6.30} &\text{96.88\%}\\
\text{5th global vertical mode} &\text{6.34} & \text{6.83}&\text{92.77\%} \\
 \hline
\end{tabular}
\end{table}

\subsection{Damage state simulation}\label{Damage state simulation}

\begin{figure}[!htb]
    \centering
    \includegraphics[width=0.7\textwidth]{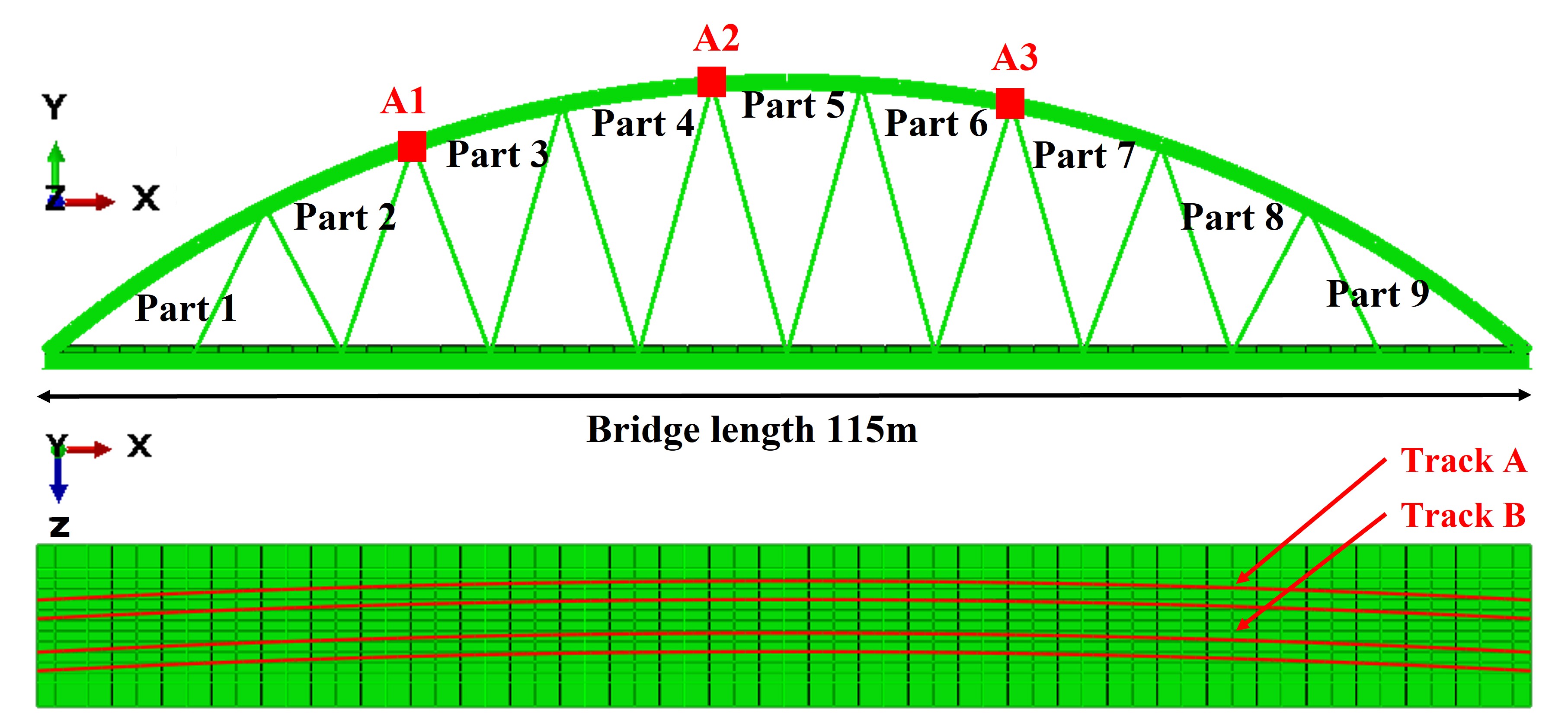}
    \caption{The locations of nine sub-parts, acceleration sensor positions (A1, A2, A3), and two tracks of KW51 bridge}
    \label{KW51-A3-sensors}
\end{figure}

Based on the validated FEM, the potential damage states on the arch of the KW51 bridge are simulated, considering different damage magnitude and locations. The magnitude of damage includes moderate damage and failure. The arch structure is segmented into nine sections, as illustrated in Figure \ref{KW51-A3-sensors}, indicating nine potential damage locations. In this study, only one damage location is evaluated at a time. The Young's modulus of the damaged part is reduced to simulate moderate damage. Moderate damage level (DL) is defined as 

\begin{equation}\label{DamageLevel}
\begin{aligned}
    DL=\left(1-\frac{E^*}{E^0} \right)\times{100\%} 
\end{aligned}
\end{equation}

\noindent where ${E^*}$ stands for the Young's modulus of the damaged part material, and ${E^0}$ stands for the Young's modulus for the intact part.
The damage level corresponds to the percentage reduction in Young's modulus. The damage levels up to 40\% are considered as moderate damage, where the damage levels of 5\%, 10\%, 15\%, and 20\% are classified as "20\%" damage class, and the damage levels of 25\%, 30\%, 35\%, and 40\% are classified as "40\%" damage class. In the future, a more developed version of this work may consider classifying damage into additional classes.

\begin{figure}[!htb]
    \centering
    \includegraphics[width=0.65\textwidth]{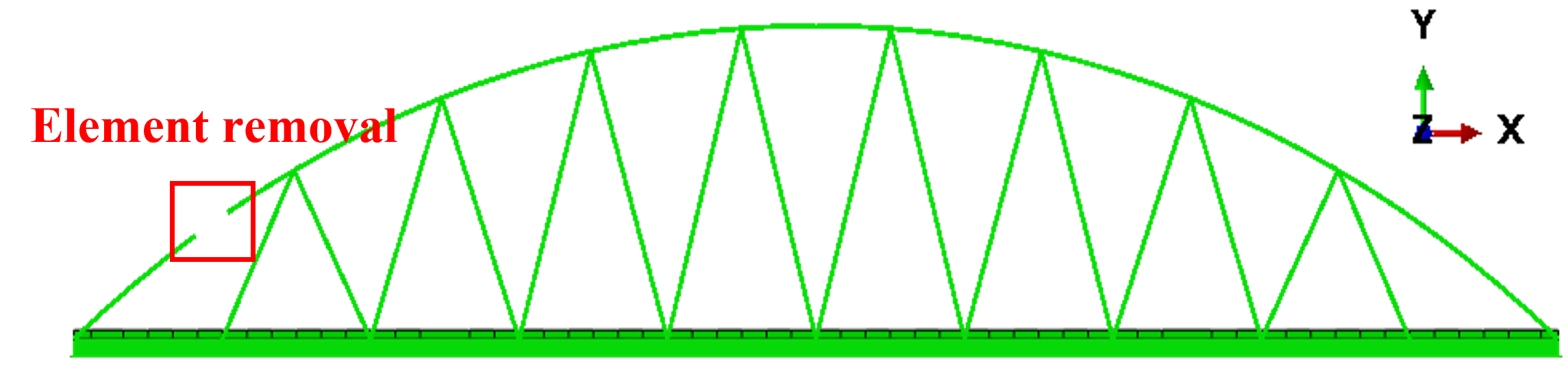}
    \caption{A schematic showing an example of the element removal approach. In this figure, the middle element on part 1 is removed.}
    \label{ElementRemovalPart1}
\end{figure}

If the stiffness reduction in the damaged section exceeds 40\%, the damage level is classified as 'Failure. To simulate failure damage, the element removal strategy is employed instead of stiffness reduction to represent disconnection in the damaged section. The element removal description in the middle of part 1 on the arch of the KW51 bridge structure is shown in Figure \ref{ElementRemovalPart1}. In each failure simulation, a single mesh element is removed from the damaged section. Nine arch parts are included in the FEM of the KW51 bridge. The number of mesh elements in these nine parts are 7, 4, 4, 4, 4, 4, 4, and 7, respectively. Therefore, a total of 42 failure simulation models are established to generate sufficient "Failure" natural frequency data.

This paper simultaneously considers the first 14 natural frequencies of the KW51 bridge and the lateral acceleration responses of the bridge arch under different damage states. The natural frequencies for the damaged states are obtained from FEM modal analysis. The dynamic acceleration response of the bridge structure is acquired through implicit analysis, applying the moving load on Track B. The acceleration responses at three positions, labeled A1, A2, and A3, are recorded as shown in Figure \ref{KW51-A3-sensors}. 

\begin{figure}[!htb]
    \centering
    \includegraphics[width=0.8\textwidth]{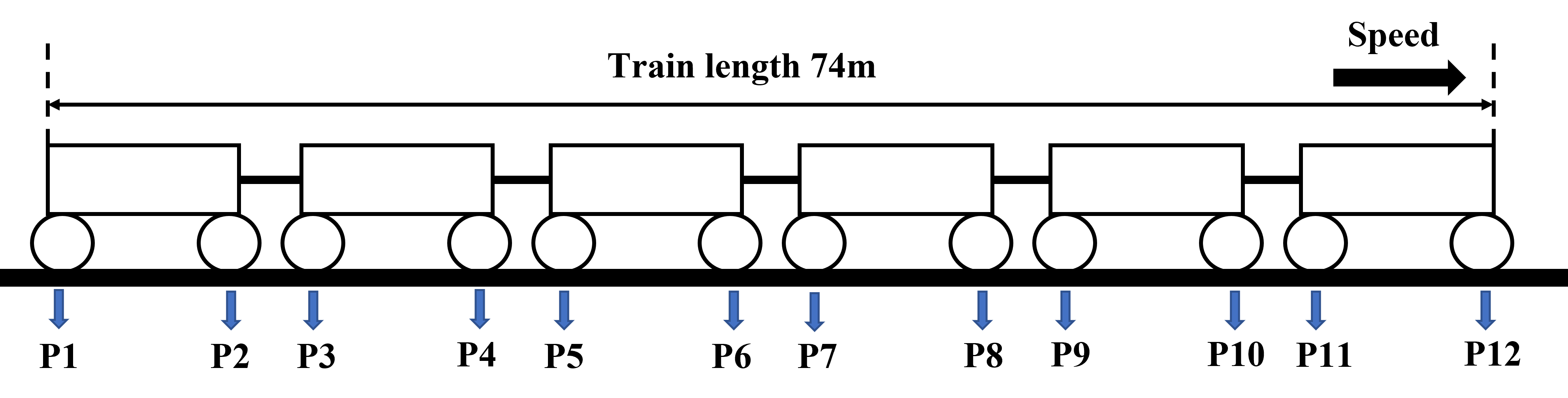}
    \caption{Diagram of the train load on KW51 bridge}
    \label{Train_load}
\end{figure}

Figure \ref{Train_load} presents a simplified diagram of the train load. The train is assumed to move from left to right on track B. Each train car is modeled with a double line load, assuming a contact length of 0.07m between each wheel and the rail. Considering six train cars, there are 12 line loads in total \cite{KW51-MSSP-stresses}. The values of line loads are defined in Table \ref{Numerical values of train loads}.
\begin{table}[h]\footnotesize
\centering
\caption{Numerical values of train loads}
\label{Numerical values of train loads}
\begin{tabular}{|c|c|}
\hline
\text{Load} & \text{Value}
\\ \hline
\text{P1}&\text{942857N/m}\\
\text{P2}&\text{938571N/m}\\
\text{P3}&\text{834285N/m}\\
\text{P4}&\text{831428N/m}\\
\text{P5}&\text{676428N/m}\\
\text{P6}&\text{675714N/m}\\
\text{P7}&\text{675714N/m}\\
\text{P8}&\text{676428N/m}\\
\text{P9}&\text{831428N/m}\\
\text{P10}&\text{834285N/m}\\
\text{P11}&\text{938571N/m}\\
\text{P12}&\text{942857N/m}\\
\hline
\end{tabular}
\end{table}

In the FEM, the line load is modeled as a simplified moving load using the Dload subroutine. This study examines six different moving speeds: 15.75 m/s, 18 m/s, 20 m/s, 21 m/s, 25 m/s, and 27 m/s. These speeds are derived from accelerator monitoring data of the KW51 bridge, recorded between October 2019 and December 2019. The time step interval in the FEM is set to 0.0012s, matching the recorded acceleration measurements. All acceleration responses are normalized based on the maximum absolute value of the acceleration, both from the FEM and the sensors data on the KW51 bridge. The normalized acceleration response avoids errors caused by unknown moving load amplitudes.

\section{Signal processing}\label{Section4-Signal processing}

Signal processing methods are crucial when unstable vibration signals are used for damage identification. These methods can quickly and effectively obtain specific signal features, such as frequency or time-frequency characteristics, making them highly sensitive to damage. This paper employs signal processing techniques such as time stacking, Fourier transform, and wavelet transform to extract these crucial features.

\subsection{Stacking}\label{Stacking}

\begin{figure}[!htb]
    \centering
    \includegraphics[width=0.9\textwidth]{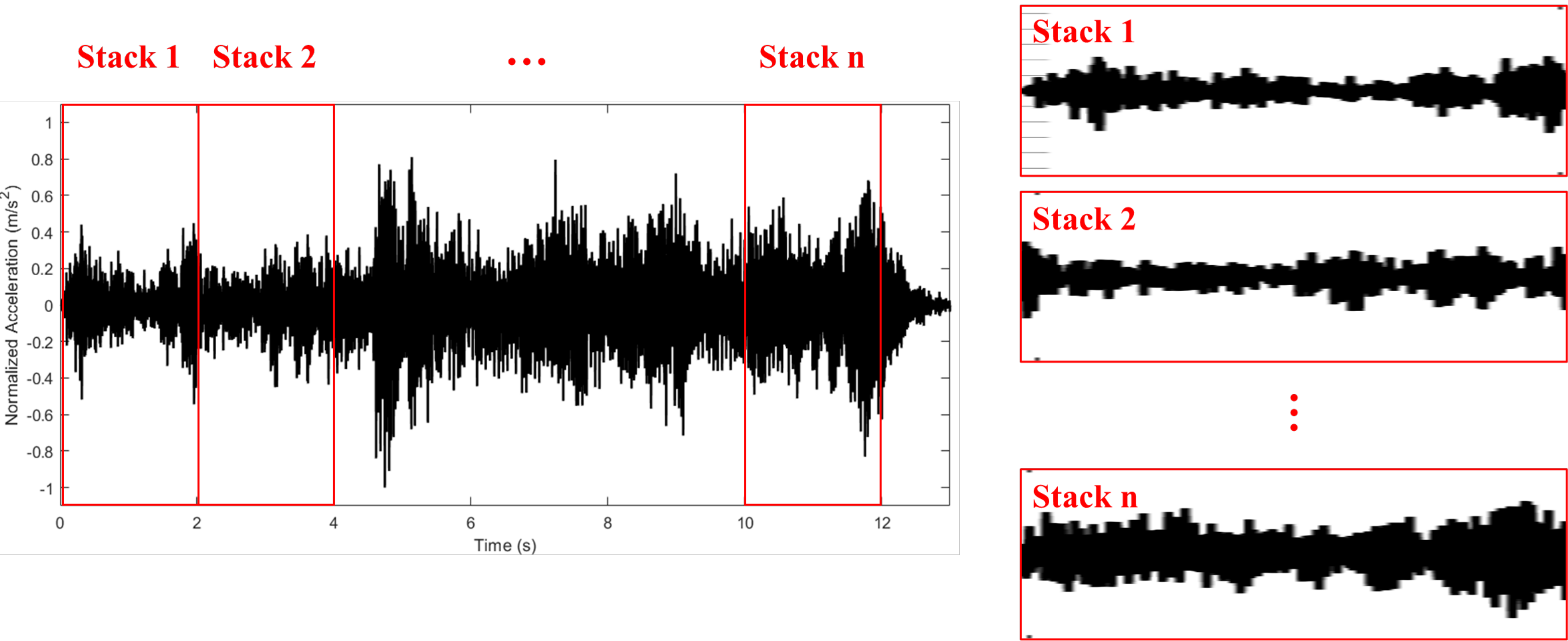}
    \caption{Stacking the time series (the stack number of n) with same length of each stacked series}
    \label{Stacked-GRU-figure}
\end{figure}

The original normalized acceleration sequence $\boldsymbol{A}=\{\boldsymbol{a}_{1}, \boldsymbol{a}_{2}, ..., \boldsymbol{a}_{t}\} $ belongs to $\boldsymbol{R}^{t\times{1}}$ is broken into multiple stacks to form a new input $\boldsymbol{A}^{s}=\{\boldsymbol{a}_{1}^{s},\boldsymbol{a}_{2}^{s}, ...,\boldsymbol{a}_{z}^{s}\}\in \boldsymbol{R}^{z\times{n}}$, where z is the new time step for each stack, and n is the total number of stacks. The number of time steps in every stack is the same \cite{Stacked-LSTM}. The stacked time series are defined in Figure \ref{Stacked-GRU-figure}. The original normalized acceleration sequence is divided into n parts and stacked together, which means converting the original single feature (acceleration) into n feature series in the temporal dimension. This stacking can significantly reduce the training time of the network \cite{Generalized-stacked-LSTM}. Although the stacked series is no longer continuous in the time dimension, it still retains a part of the structural vibration characteristics. The smaller the number of stacks, the more complete the structural vibration characteristics retained by the stacked data. Therefore, selecting an appropriate number of stacks is important to balance computational efficiency and accuracy.

\subsection{Fourier transform}\label{Fourier transform}

The frequency domain characteristics of the acceleration response are critical for analyzing the structural state. Structural damage can cause the acceleration response to shift in a certain frequency band. Therefore, the Fourier transform is used to obtain the frequency domain characteristics of acceleration. Fourier transform is defined as \cite{Fourier-Transform}:

\begin{equation}\label{Section4-Fourier-transform}
\begin{aligned}
    \boldsymbol{F}(\omega) = \int_{-\infty}^{+\infty} \boldsymbol{A}(t)e^{-i\omega t}dt
\end{aligned}
\end{equation}

\noindent where $\boldsymbol{A}(t)$ is the acceleration function in time domain. To unify the frequency series, the frequency sequence obtained by the Fourier transform is to be normalized according to the maximum module value, which is defined as:

\begin{equation}\label{Normalized F}
\begin{aligned}
    \boldsymbol{F}^n(\omega) = \frac{\lvert {\boldsymbol{F}(\omega)} \rvert}{\max \lvert {\boldsymbol{F}(\omega)} \rvert}
\end{aligned}
\end{equation}

\begin{figure}[!htb]
    \centering
    \includegraphics[width=0.8\textwidth]{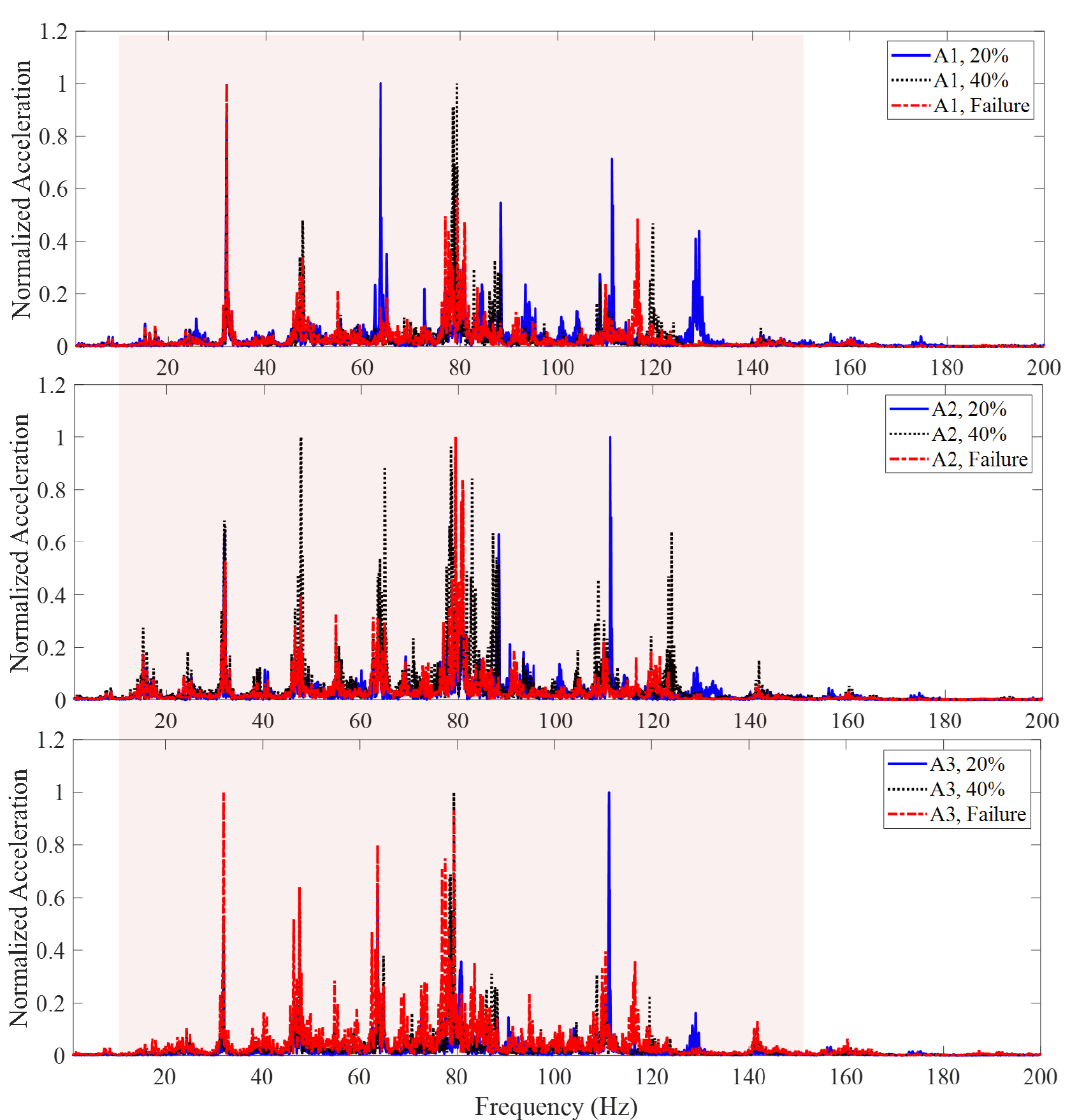}
    \caption{The frequency curves of accelerations from A1, A2 and A3, with damage levels of 20\%, 40\%, and failure}
    \label{kNN-Acc-Frequency-A123-15ms}
\end{figure}

The normalized frequency characteristic series is obtained from the acceleration time history. The sampling frequency of acceleration is 833Hz, but the frequency characteristics are concentrated below 150Hz. In order to clearly explain the difference in frequency characteristics, a local graph with a frequency range from 1Hz to 200Hz is shown in Figure \ref{kNN-Acc-Frequency-A123-15ms}. The frequency sequences within the bandwidth of 10Hz-150Hz (pink shade in Figure \ref{kNN-Acc-Frequency-A123-15ms}) are considered because the frequency domain values outside this bandwidth are negligible. This figure shows the acceleration frequency domain characteristic curves at three positions A1, A2, and A3, for the moving velocity of 15.75m/s. It compares the frequency characteristics of three different damage levels (20\%, 40\%, and Failure) in Part 1. It is found that the curves of the three damage levels are consistent in the frequency domain below 60Hz, and their differences only lie in the amplitudes. Within the frequency band of 60Hz to 150Hz, their frequency characteristics differed. Compared with the curve of 20\% damage, the peaks of 40\% and failure curves are offset and even multi-peaked in frequency direction. At the same time, the amplitude of the peaks shows an obvious difference. The normalized Fourier transform extracts the frequency characteristic differences under the different damage levels. The frequency feature values of A1, A2, and A3 are merged to improve the robustness of damage identification. 

\subsection{Wavelet transform}\label{Section4-Wavelet ransform}

A wavelet basis is a mathematical function used to decompose a signal into different frequency bands and analyze each band with a resolution that matches its scale \cite{wavelet-1998}. The continuous wavelet transform (CWT) \cite{CWT-2010} provides an over-complete representation of the vibration signal by continuously varying the translation and scaling parameters. This approach decomposes the signal into frequency bands and captures local temporal information through correlation resolution. For the acceleration-time function $\boldsymbol{A}(t) $, the CWT is expressed by the following integral \cite{wavelet-AE-CNN}:

\begin{equation}\label{CWT}
\begin{aligned}
    \boldsymbol{W}(a,b) = \int_{-\infty}^{\infty}\boldsymbol{A}(t) \frac{1}{\sqrt{a}}\psi\left( \frac{t-b}{a}\right) dt
\end{aligned}
\end{equation}

\noindent where \( a \) and \( b \) are the scaling and translation parameters, respectively, and $\boldsymbol{\Psi}(t) $ is the mother wavelet function.
The position of the wavelet in the time domain is denoted by \( b \), and its position in the frequency domain is given by \( a \). Therefore, the CWT maps the original series into a function of \( b \) and \( a \), providing simultaneous information on both time and frequency. In this paper, the Morlet wavelet is employed, which can be expressed as follows:

\begin{equation}\label{Morlet wavelet}
\begin{aligned}
    \boldsymbol{\Psi}(t) = \pi^{-\frac{1}{4}}e^{i\omega_{0} t}e^{-\frac{t^2}{2}}
\end{aligned}
\end{equation}

\begin{figure}[!htb]
    \centering
    \includegraphics[width=0.88\textwidth]{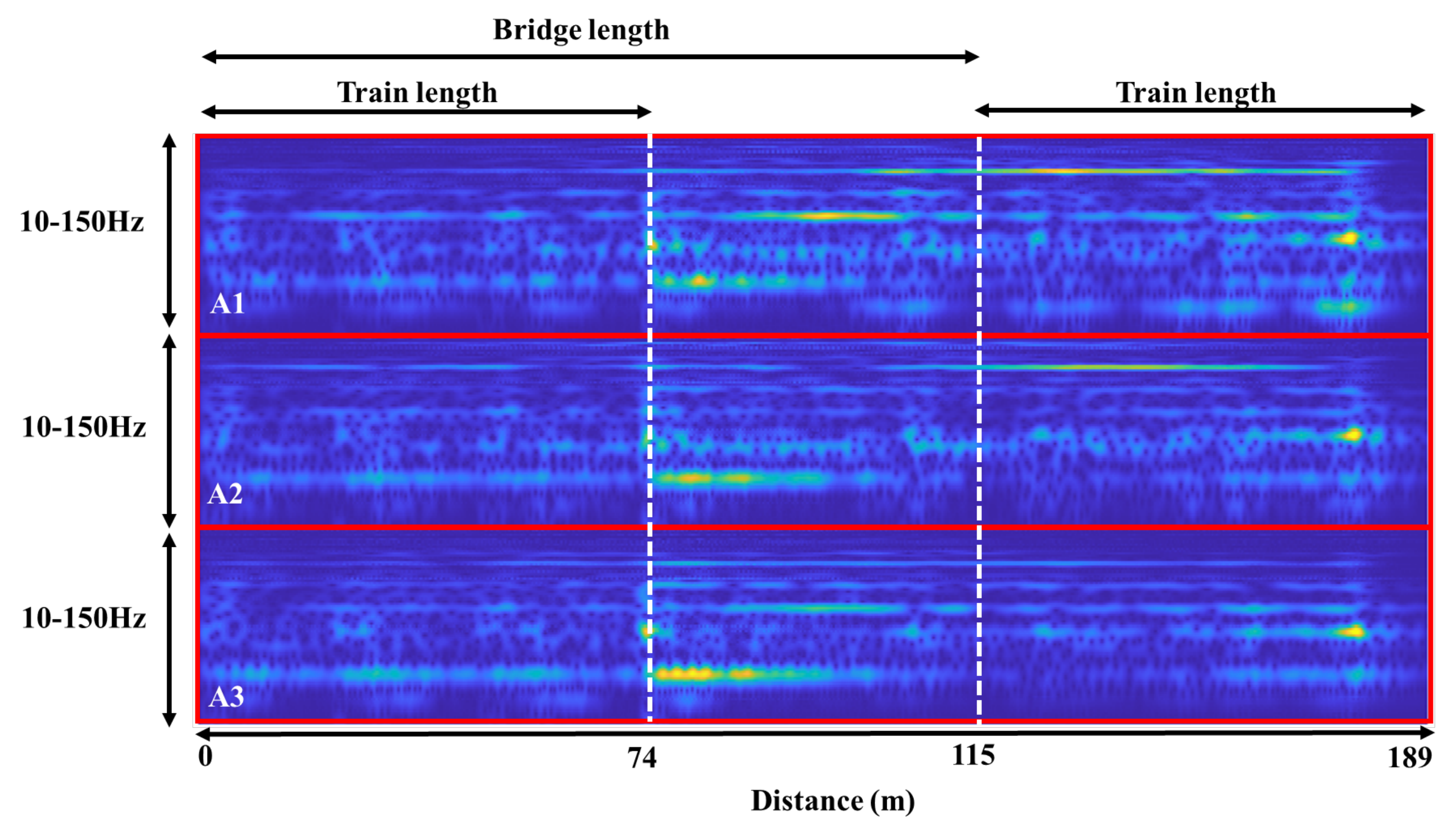}
    \caption{Distance-frequency image obtained from the wavelet transform of A1, A2, and A3 sensors.}
    \label{wavelet-A3}
\end{figure}

Morlet wavelet is particularly used for analyzing non-stationary signals where time and frequency location identification is crucial. The time-frequency characteristics of acceleration signals are visualized after the CWT. For better visualization, this study converts the time on the x-axis to distance, obtained by multiplying time by train speed, which is defined as:

\begin{equation}\label{Time to displacement}
\begin{aligned}
    \boldsymbol{u}(t) = \int_{0}^{t}\boldsymbol{v}(t)dt
\end{aligned}
\end{equation}

\noindent where $\boldsymbol{v}(t)$ is the velocity of moving load. In this study, the moving speed of the train is assumed to be constant, which can be estimated from acceleration. $\boldsymbol{u}(t)$ is the distance-time function. In this way, the x-axis represents distance, and the y-axis represents frequency. The distance range is 0m to 189m, and a frequency band from 10Hz to 150Hz is considered. For comprehensive analysis, three time-frequency images, corresponding to sensors A1, A2, and A3, are combined into a single image, as shown in Figure \ref{wavelet-A3}. This approach allows for a detailed examination of the dynamic responses captured by multiple sensors simultaneously.

Figure \ref{wavelet-A3} shows the entire process from a train approach to the departure of the bridge. The process can be divided into three stages along the x-axis: the train approaching, the train fully on the bridge, and the train departing. Initially, as the train enters, the acceleration response on the bridge arch remains small. Once the train is fully on the bridge, the acceleration response increases. As the train exits, the response decreases, but a brief fluctuation occurs due to the step-like reduction of the moving load. From the y-axis, the frequency characteristics of the acceleration response are primarily concentrated below 150Hz. Notably, when the trains are fully on the bridge, the frequency characteristics of A1, A2, and A3 exhibit clear differences. These variations are evident in the position, size, and brightness of the highlighted areas on the image. These distinct features displayed in the image will be used for damage location identification.

\section{Machine learning approaches}\label{Section5-ML}

This section introduces the ML approaches deployed in this paper. In this work, the damage identification problem is transformed into a classification problem. 
The following specific input types are considered: frequency domain features, time series data, and three-dimensional images. The GRU method is applied to identify the existence of damage \cite{GRU,Stacked-LSTM,Compare-LSTM-GRU} due to its distinct advantages in handling time series, including high accuracy and time efficiency  \cite{DI-GRU,GRU-Prediction}. While the kNN classifier is used for feature-based input \cite{k-KNN,KNN,kNN-1989,kNN-high} as kNN is a well-established classification method \cite{apply-KNN,Frequency-KNN-mssp}.
The CNN algorithm is adopted for image-based input \cite{wavelet-AE-CNN,simpleCNN} to localize the damage. The images are derived from wavelet transforms, containing sensitive information of the damage location. CNNs are well known to be a powerful setup for machine learning of visual data \cite{CNN-Concrete-Building,time-fre-CNN}.

\subsection{Stacked GRU}\label{Stack GRU}
\begin{figure}[!htb]
    \centering
    \includegraphics[width=0.8\textwidth]{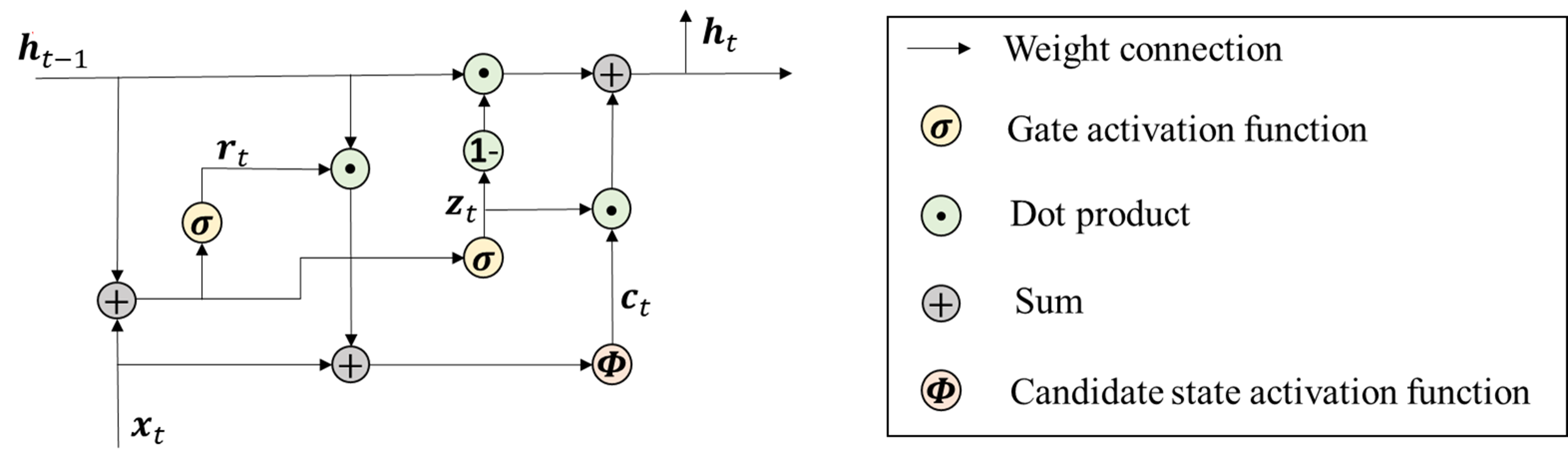}
    \caption{Diagram of a GRU cell}
    \label{GRU cell}
\end{figure}

GRU is an efficient method for dealing with time series data. A GRU cell comprises two gates: the reset gate and the update gate, as shown in Figure \ref{GRU cell}. $\boldsymbol{h}(t)$ denotes the hidden output at time t, while $\boldsymbol{h}(t-1)$ represents the hidden output at the preceding time step. The reset gate $\boldsymbol{r}(t)$ determines whether all or part of $\boldsymbol{h}(t-1)$ (the previous step) is considered. While the update gate $\boldsymbol{z}(t)$ controls the extent of the update to $\boldsymbol{h}(t)$ (the current step) based on the state of the candidate layer $\boldsymbol{c}(t)$ \cite{GRU}. The gate activation function $\boldsymbol{\sigma}$ and the candidate state activation function $\boldsymbol{\Phi}$ are expressed as:

\begin{equation}\label{Sigma gate}
\begin{aligned}
    \boldsymbol{\sigma}(t) = {\frac{1}{1+e^{-t}}}
\end{aligned}
\end{equation}

\begin{equation}\label{Phi gate}
\begin{aligned}
    \boldsymbol{\Phi}(t) = {\frac{e^{t}-e^{-t}}{e^{t}+e^{-t}}}
\end{aligned}
\end{equation}

For a time-series sample set $\boldsymbol{x}$(t), the calculation procedures of GRU are

\begin{equation}\label{z(t)}
\begin{aligned}
    \boldsymbol{z}(t) = \boldsymbol{\sigma}(\boldsymbol{V}_{xz}\boldsymbol{x}(t)+\boldsymbol{U}_{hz}\boldsymbol{h}(t-1)+\boldsymbol{b}_{z})
\end{aligned}
\end{equation}

\begin{equation}\label{r(t)}
\begin{aligned}
    \boldsymbol{r}(t) = \boldsymbol{\sigma}(\boldsymbol{V}_{xr}\boldsymbol{x}(t)+\boldsymbol{U}_{hr}\boldsymbol{h}(t-1)+\boldsymbol{b}_{r})
\end{aligned}
\end{equation}

\begin{equation}\label{c(t)}
\begin{aligned}
    \boldsymbol{c}(t) = \boldsymbol{\Phi}(\boldsymbol{V}_{xc}\boldsymbol{x}(t)+\boldsymbol{U}_{hc}(\boldsymbol{r}(t)\bigodot\boldsymbol{h}(t-1))+\boldsymbol{b}_{c})
\end{aligned}
\end{equation}

\begin{equation}\label{h(t)}
\begin{aligned}
    \boldsymbol{h}(t) = (1-\boldsymbol{z}(t))\bigodot\boldsymbol{h}(t-1)+\boldsymbol{z}(t)\bigodot\boldsymbol{c}(t)
\end{aligned}
\end{equation}
\noindent where $\boldsymbol{V}_{xz}$, $\boldsymbol{V}_{xr}$, and $\boldsymbol{V}_{xc}$ denote the weights linking the input layer to the update gate, reset gate, and candidate layer of the GRU network, respectively. Meanwhile, $\boldsymbol{U}_{hz}$, $\boldsymbol{U}_{hr}$, and $\boldsymbol{U}_{hc}$ represent the self-connection weights between the current time t and the previous time t-1, facilitating the network ability to capture temporal dependencies. Additionally, $\boldsymbol{b}_{z}$, $\boldsymbol{b}_{r}$, and $\boldsymbol{b}_{c}$ stand for the biases associated with the update gate, reset gate, and candidate layer within the GRU unit, respectively. These biases play a critical role in determining the threshold for gate activation and thus influence the flow of information within the GRU unit. 

\begin{figure}[!htb]
    \centering
    \includegraphics[width=0.7\textwidth]{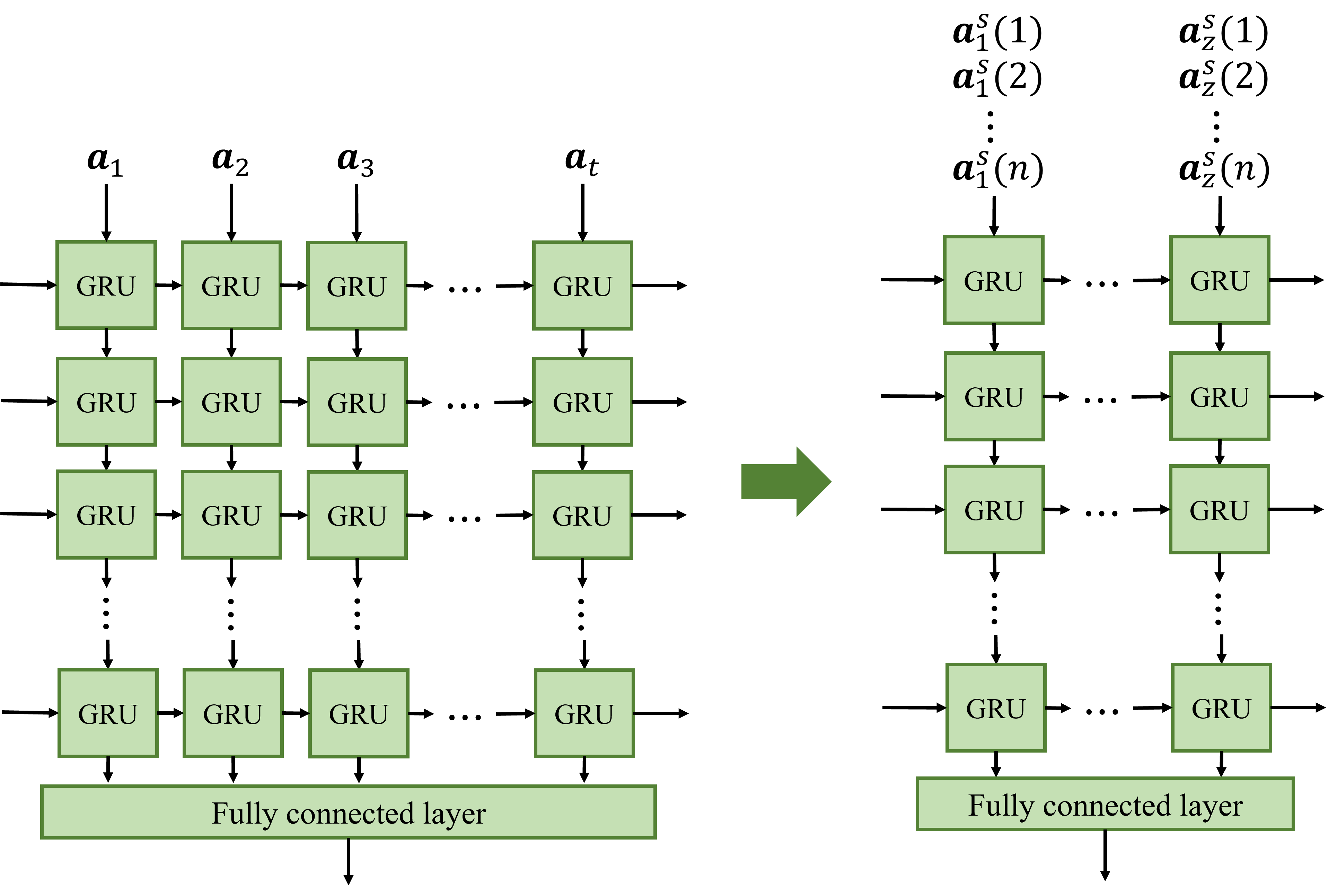}
    \caption{The diagram of stacked GRU network}
    \label{StackedStackedGRU}
\end{figure}

Based on the stacked LSTM classifier proposed by Ahmed et al. \cite{Stacked-LSTM}, a stacked GRU method is adopted in this paper. 
The stacked GRU method reduces the complexity of the GRU network by using a stacked input time series. The figure illustrates the principle of optimizing the GRU network by stacking sequences. When the original time series is used as the input, more GRU cells are required to capture the temporal correlations. At each time step, the chunks of all stacks (single value of each stack) are fed into the GRU simultaneously, and each stack is regarded as a new input feature of the network. By selecting an appropriate number of stacks, the length of the stacked sequence is shortened, reducing the number of required GRU units. The diagram of the stacked GRU network is shown in Figure \ref{StackedStackedGRU}. The choice of the number of stacks depends on the time signal correlation. For weak correlations, such as the stable periodic signal, a larger number of stacks improves training efficiency, while for strong correlations such as the non-stationary signal, the number of stacks needs careful consideration. The discussion on the number of stacks is provided in Section \ref{Acceleration damage magnitude}.

\subsection{kNN algorithm}\label{kNN}

kNN algorithm is an effective machine learning algorithm that relies on the supervised learning technique. The kNN classifier operates through three straightforward steps:
\begin{enumerate}
    \item \textbf{Calculate Distances:} Determine the distance between the prediction sample and each labeled sample in the dataset.
    \item \textbf{Select Neighbors:} Identify the k-nearest neighbors based on the smallest distances calculated.
    \item \textbf{Assign Label:} Assign the predicted label to the new sample based on the majority label among the k-nearest neighbors.
\end{enumerate}

\begin{figure}[!htb]
    \centering
  \includegraphics[width=0.4\textwidth]{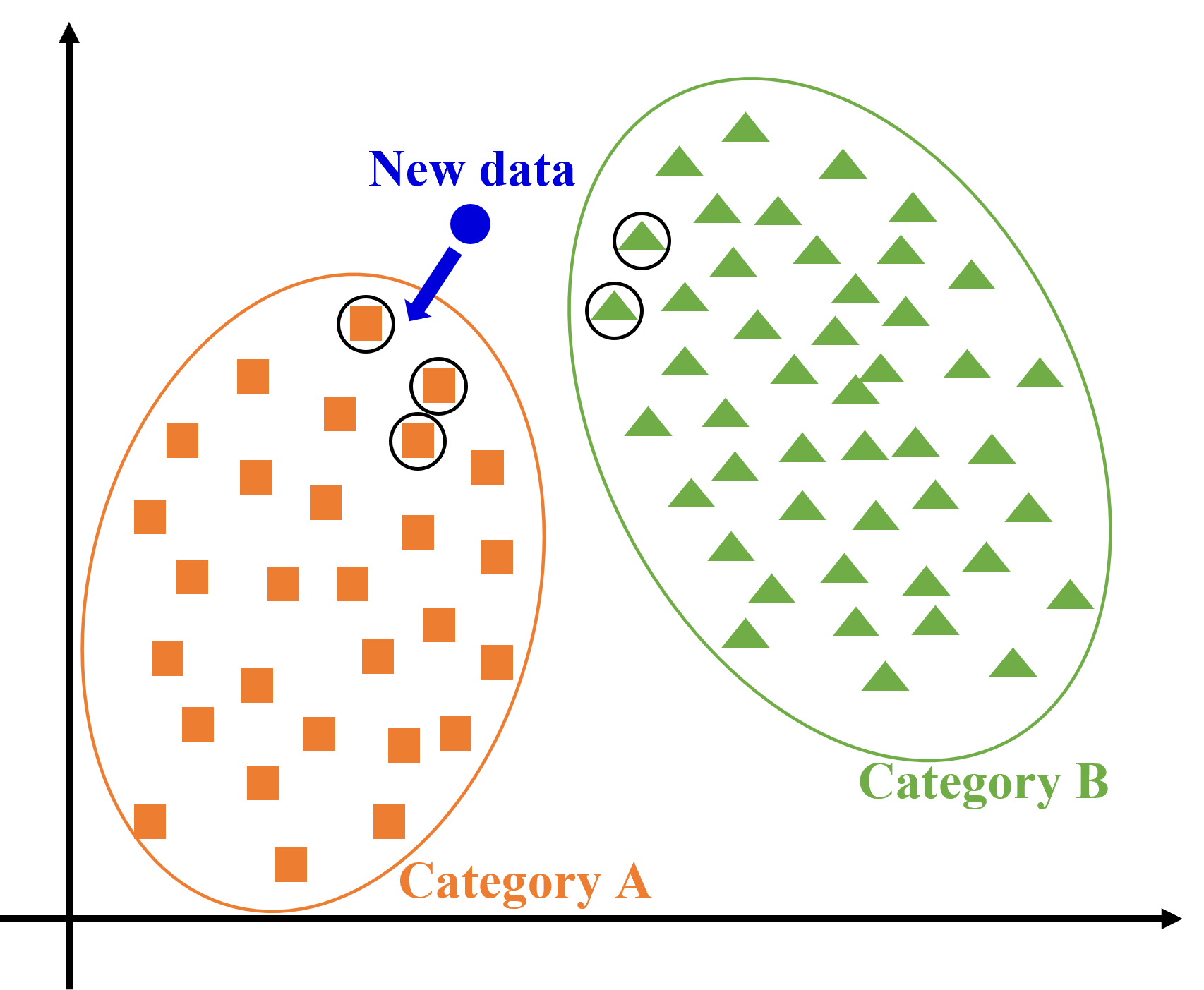}
    \caption{A kNN classifier example of binary classification with a k value of 5}
    \label{kNN classify strategy}
\end{figure}

Following these steps, the kNN classifier effectively predicts the category of new data points based on the characteristics of existing labeled data. Figure \ref{kNN classify strategy} shows a kNN classifier example of binary classification with a k value of 5 \cite{kNN-book}.

Various distance definitions that can be applied to the kNN algorithm. The most classical distance is the Euclidean distance. There are also Chebyshev distance, Minkowski distance, and so on. Euclidean distance (D$^{E}$), Chebyshev distance (D$^{C}$), and Minkowski distance (D$^{M}$) are defined as:

\begin{equation}\label{Euclidean distance}
\begin{aligned}
    D^{E}= \sqrt{\frac{1}{N}\sum_{k=1}^{N}(x_{ik}-x_{jk})^{2}}
\end{aligned}
\end{equation}
\begin{equation}\label{Chebyshev distance}
\begin{aligned}
    D^{C}= \max_{k}(\lvert{x_{ik}-x_{jk}}\rvert)
\end{aligned}
\end{equation}
\begin{equation}\label{Minkowski distance}
\begin{aligned}
    D^{M}= \left( {\frac{1}{N}\sum_{k=1}^{N}(\lvert{x_{ik}-x_{jk}}\rvert)^{p}}\right)^{\frac{1}{p}} 
\end{aligned}
\end{equation}

\noindent where x$_{i}$ and x$_{j}$ represent the features of label i and label j. N is the length of features. The exponent p is defined within the interval [0.5, 3] for Minkowski distance, which is applied in \ref{Minkowski distance}.

Meanwhile, the accuracy of the kNN algorithm is closely related to the value of k. Generally, a larger k value tends to result in higher accuracy, but it can also lead to the appearance of outliers. Therefore, an automatic algorithm is adopted by looping the k value from 1 to N/2, using multiple distance equations. The parameter values of p and k are determined by minimizing the cross-validation loss error. 

\subsection{CNN}\label{CNN}
CNN offers significant advantages for image classification problems \cite{wavelet-AE-CNN,time-fre-CNN}. This paper leverages the wavelet transform, which excels at simultaneously capturing time-frequency domain information from vibration signals, to obtain the time-frequency domain feature images of the bridge's acceleration response. Subsequently, bridge damage identification is achieved using the CNN classifier.

\begin{figure}[!htb]
    \centering
    \includegraphics[width=0.6\textwidth]{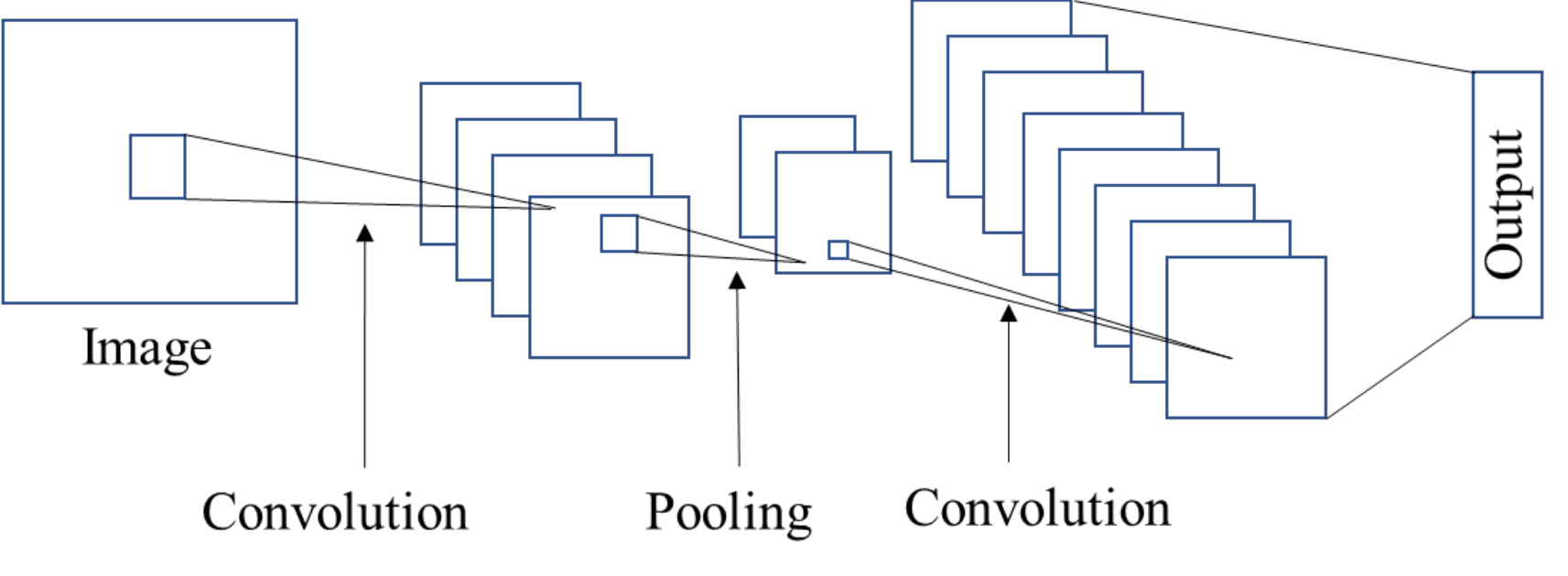}
    \caption{An example of CNN network}
    \label{CNN-0}
\end{figure}

A CNN network consists of several convolution layers with activation functions and pooling layers. Figure \ref{CNN-0} shows a simple CNN network, containing the resized image input, two convolution layers, one pooling layer, and an output layer. Image pixels can be directly used as input to standard feed-forward neural networks for image classification problems. Convolution layers are the key components of CNNs. In image classification tasks, one or more matrices/channels serve as the input to the convolution layer, and multiple matrices are produced as the output. The number of input and output matrices can be different. The process to compute a single output matrix is defined as follows \cite{CNN-Convolution}:
\begin{equation}\label{Convolution}
\begin{aligned}
    \boldsymbol{M}_{j}=f(\sum_{i=1}^{N} \boldsymbol{I}_{i}\times \boldsymbol{K}_{i,j}+\boldsymbol{B}_{j})
\end{aligned}
\end{equation}

\noindent where $\boldsymbol{I}_{i} $ is input matrix, which is convoluted with the kernel matrix $\boldsymbol{K}_{i,j} $. Then the sum of all convoluted matrices is computed and a bias value $\boldsymbol{B}_{j} $ is added to each element of the resulting matrix. Finally a non-linear activation function f is applied to each element of the previous matrix to produce one output matrix $\boldsymbol{M}_{j} $.

\subsection{Damage identification strategy for the KW51 bridge}\label{SHM-FlowChart}

 The previous subsection introduces the machine learning approaches for damage identification. This work considers both modal analysis-based and acceleration-based strategies. The damage identification strategy based on modal analysis uses the natural frequency values as input. Since the natural frequency data has a simple structure and is a damage-sensitive feature, kNN classifiers are employed to categorize the natural frequency dataset for damage identification. However, due to the complexity and time-consuming nature of modal analysis, the damage identification step based on dynamic analysis is necessary in certain cases. This method utilizes various damage-sensitive features extracted from acceleration signals. To estimate the damage magnitude, stacked time series and frequency sequences serve as inputs for stacked GRU and kNN algorithms, respectively. Additionally, time-frequency images of acceleration, which highlight damage location-sensitive features, are used with a CNN algorithm for damage location identification. In the following of this paper, the damage identification methods based on modal analysis and acceleration are discussed in Section \ref{Section6-frequency} and Section \ref{Section7-acceleration}, respectively.

\section{Damage identification based on modal analysis of CMLDI method} \label{Section6-frequency}

The natural frequency is an inherent characteristic of the structure. When structural damage occurs, the stiffness changes or the failure of the structure will cause significant variations in the natural frequency. This provides a theoretical basis for using natural frequency to identify damage in the KW51 bridge. This paper employs a validated FEM to generate natural frequency data under various damage states for the KW51 bridge structure. These data are combined with natural frequency measurements obtained through long-term structural health monitoring to achieve damage identification and location identification. The "Intact" samples come from measured natural frequency values and FEM modal analysis, while the "Damaged" data are all generated from the FEM.

\begin{figure}[!htb]
    \centering
    \includegraphics[width=0.65\textwidth]{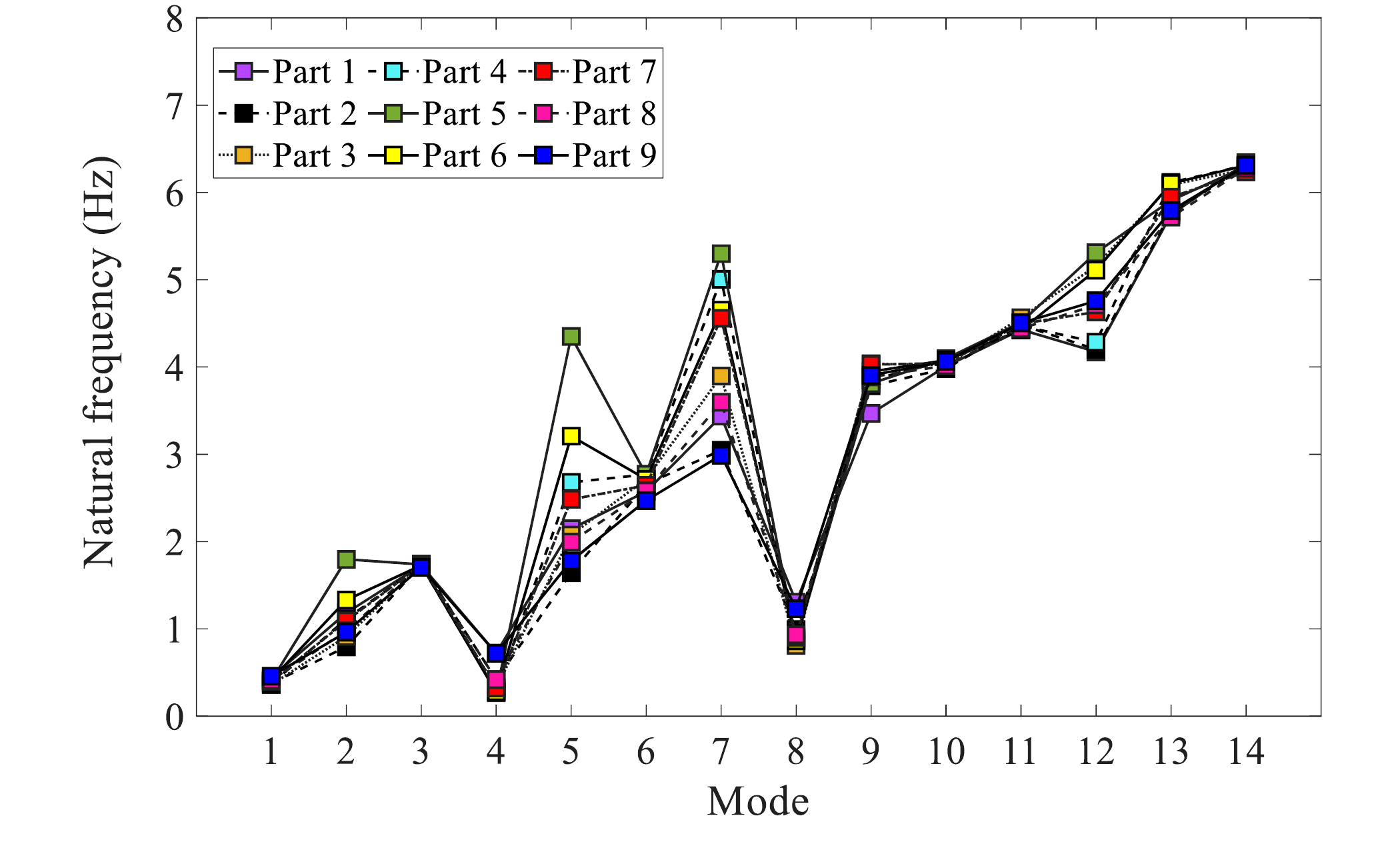}
    \caption{FEM natural frequencies of KW51 bridge with nine failure states }
    \label{Failure frequency}
\end{figure}

Figure \ref{Failure frequency} shows the FEM natural frequencies of the KW51 bridge under nine partial failure states. The natural frequency of part 5, located in the middle of the bridge arch, has the most significant variation based on damage. When structural damage occurs on the arch, the arch lateral modal frequencies and global modal frequencies change significantly, especially the 2nd, 3rd, and 4th order arch lateral modes and the 3rd and 4th order global modes.

\begin{figure}[!htb]
    \centering
    \includegraphics[width=0.65\textwidth]{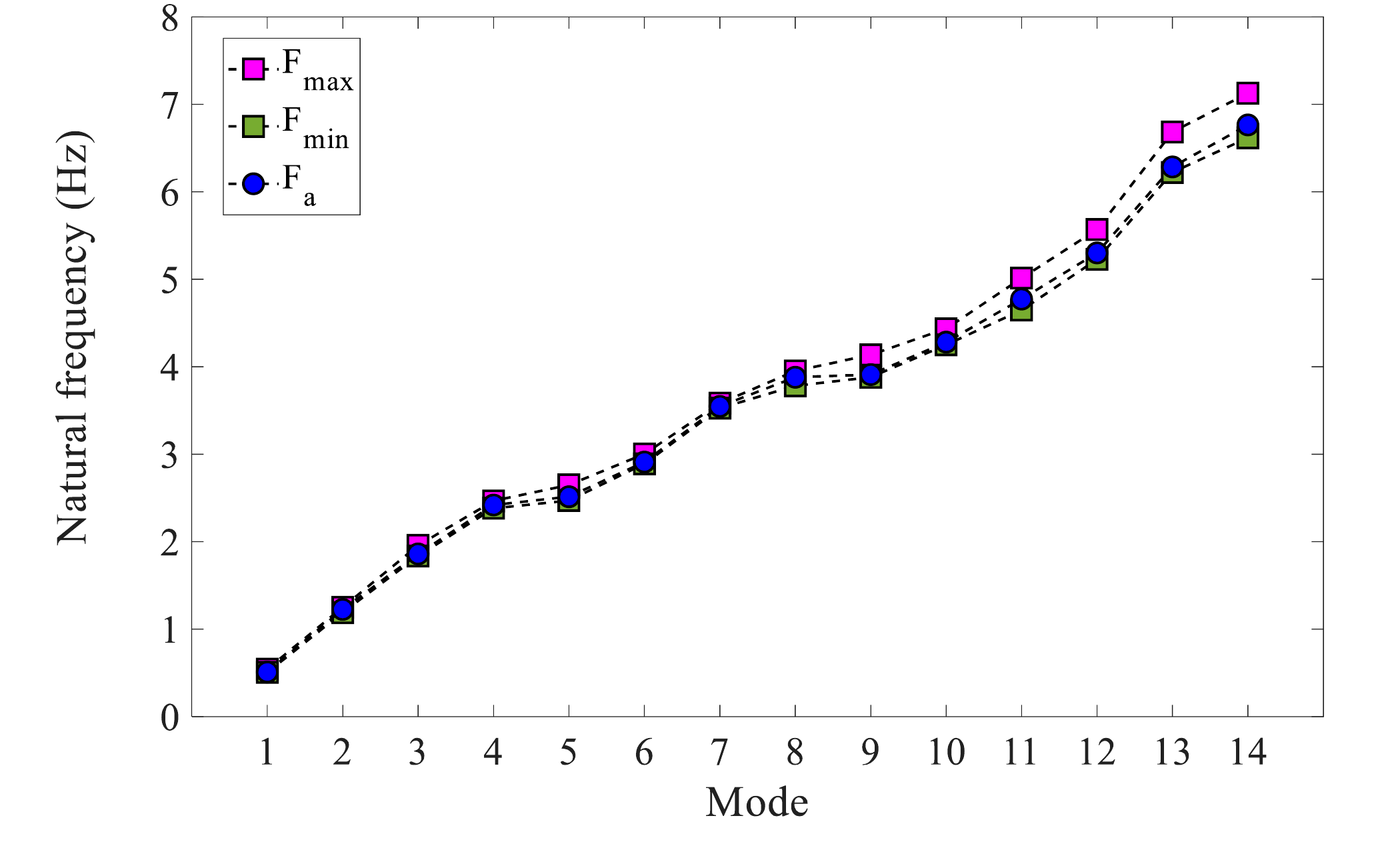}
    \caption{The average, maximum, and minimum values of measured natural frequencies}
    \label{measured-Max-Min-frequency}
\end{figure}

The measured natural frequency data is used for the "Intact" dataset. As Figure \ref{measured-Max-Min-frequency} shows, the measured frequency data are the results of long-term health monitoring of the KW51 bridge. This paper selects 3000 sets of measurement results and calculates the maximum, minimum and average values of the first 14 natural frequencies, which are presented in this figure. The measurement error is defined as:
\begin{equation}\label{Measured-fre-error-max}
\begin{aligned}
    \boldsymbol{E}_{max} = {\frac{\boldsymbol{F} _{max}-\boldsymbol{F} _a}{ \boldsymbol{F} _a}}\times{100\%}
\end{aligned}
\end{equation}
\begin{equation}\label{Measured-fre-error-min}
\begin{aligned}
    \boldsymbol{E}_{min} = {\frac{\boldsymbol{F} _{min}-\boldsymbol{F} _a}{ \boldsymbol{F} _a}}\times{100\%}
\end{aligned}
\end{equation}

\noindent where $\boldsymbol{F} _a$ is the average measured frequency, $\boldsymbol{F} _{max}$ and $\boldsymbol{F} _{min}$ are respectively the maximum and minimum measured frequencies. $\boldsymbol{E}_{max}$ and $\boldsymbol{E}_{min}$ stand for the Max. measured frequency and the Min. measured frequency.

\begin{figure}[!htb]
    \centering
    \includegraphics[width=0.65\textwidth]{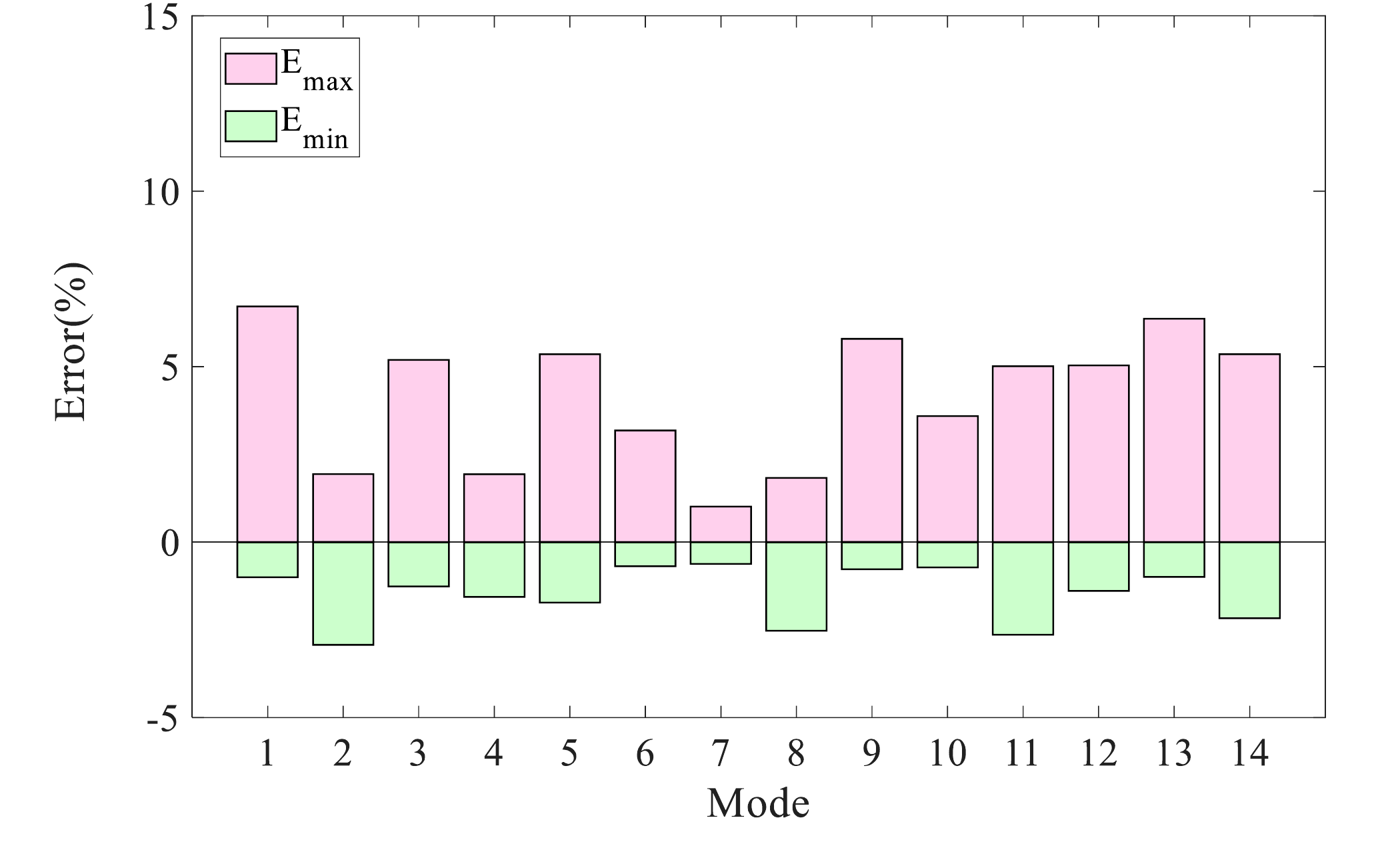}
    \caption{The maximum and minimum errors of measured natural frequencies}
    \label{measured-Max-Min-frequency-Err}
\end{figure}

As shown in Figure \ref{measured-Max-Min-frequency-Err}, the measured natural frequency contains environmental noise,  resulting in a maximum error of 6.7\% and an average ambient noise level of 3\%. Therefore, the natural frequency obtained from the FEM is utilized with two scenarios: 1- Case 1, which excludes natural frequency noise, and 2- Case 2, which includes frequency uncertainty within a 7\% range. Case 1 contains 114 datasets of natural frequency in damage states, with 36 sets labeled as 20\% and 40\%, respectively, and 42 sets labeled as a failure. To balance the dataset, 39 sets of measured natural frequencies and 1 set of natural frequencies from FEM modal analysis are used as the "Intact" labeled data. Since Case 2 considers a frequency error of 0\%-7\% at 1\% intervals, the damage dataset is expanded to 912 sets, including 288 sets labeled as 20\% and 40\%, respectively, and 336 sets for failure. In this case, 279 sets of measurement data and 1 set of FEM data are used to label the intact state. The description for the total samples is listed in Table \ref{Total samples Case 1} and \ref{Total samples Case 2}.

\begin{table}[h]\footnotesize
\centering
\caption{The description for the samples of Case 1 (without considering the noise into FEM natural frequency)}
\label{Total samples Case 1}
\begin{tabular}{|c|c|c|c|c|}
\hline
 \text{Noise level} &\text{Description} & \text{Label}&\text{Source}&\text{Samples}\\ \hline
\text{3\%}&\text{Intact}&\text{Intact}&\text{Measured, FEM}&\text{40}\\
\text{0\%}&\text{Damage levels: 5\%, 10\%, 15\%, 20\%}&\text{20\%}&\text{FEM}&\text{36}\\
\text{0\%}&\text{Damage levels: 25\%, 30\%, 35\%, 40\%}&\text{40\%}&\text{FEM}&\text{36}\\
\text{0\%}&\text{Failure}&\text{Failure}&\text{FEM}&\text{42}\\
\hline
\end{tabular}
\end{table}

\begin{table}[h]\footnotesize
\centering
\caption{The description for the samples of Case 2 (considering the noise into FEM natural frequency)}
\label{Total samples Case 2}
\begin{tabular}{|c|c|c|c|c|}
\hline
 \text{Noise level} &\text{Description} & \text{Label}&\text{Source}&\text{Samples}\\ \hline
\text{3\%}&\text{Intact}&\text{Intact}&\text{Measured, FEM}&\text{280}\\
\text{0\%-7\%}&\text{Damage levels: 5\%, 10\%, 15\%, 20\%}&\text{20\%}&\text{FEM}&\text{288}\\
\text{0\%-7\%}&\text{Damage levels: 25\%, 30\%, 35\%, 40\%}&\text{40\%}&\text{FEM}&\text{288}\\
\text{0\%-7\%}&\text{Failure}&\text{Failure}&\text{FEM}&\text{336}\\
\hline
\end{tabular}
\end{table}

\subsection{Damage identification by kNN}\label{DamageIdentification-kNN}

\begin{figure}[!htb]
    \centering
    \includegraphics[width=0.85\textwidth]{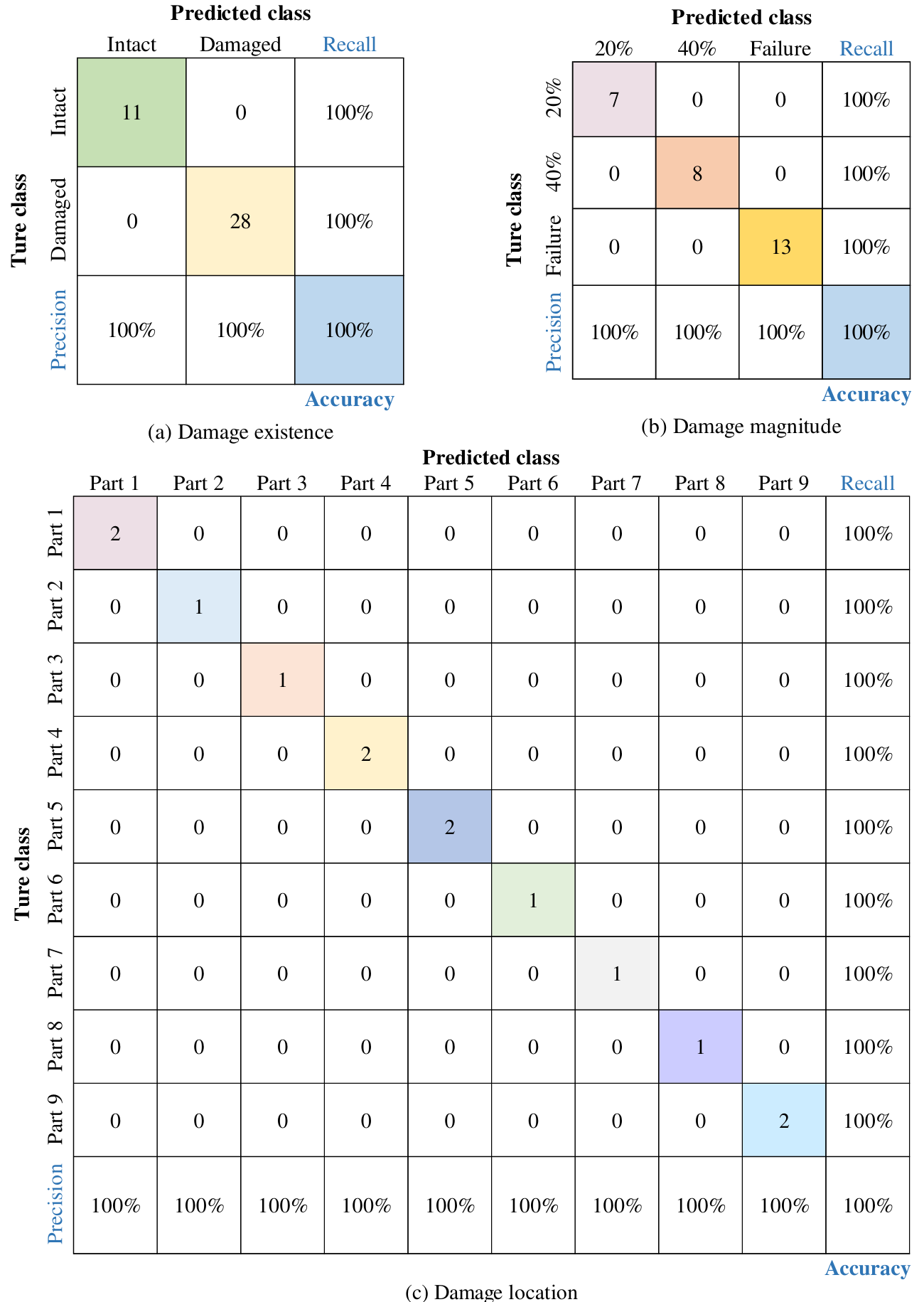}
    \caption{The damage identification results of kNN based on modal analysis under Case 1 (without considering the noise into FEM natural frequency): (a) Damage existence, (b) Damage magnitude, (c) Damage location}
    \label{kNN-classification-0Err}
\end{figure}

\begin{figure}[!htb]
    \centering
    \includegraphics[width=0.85\textwidth]{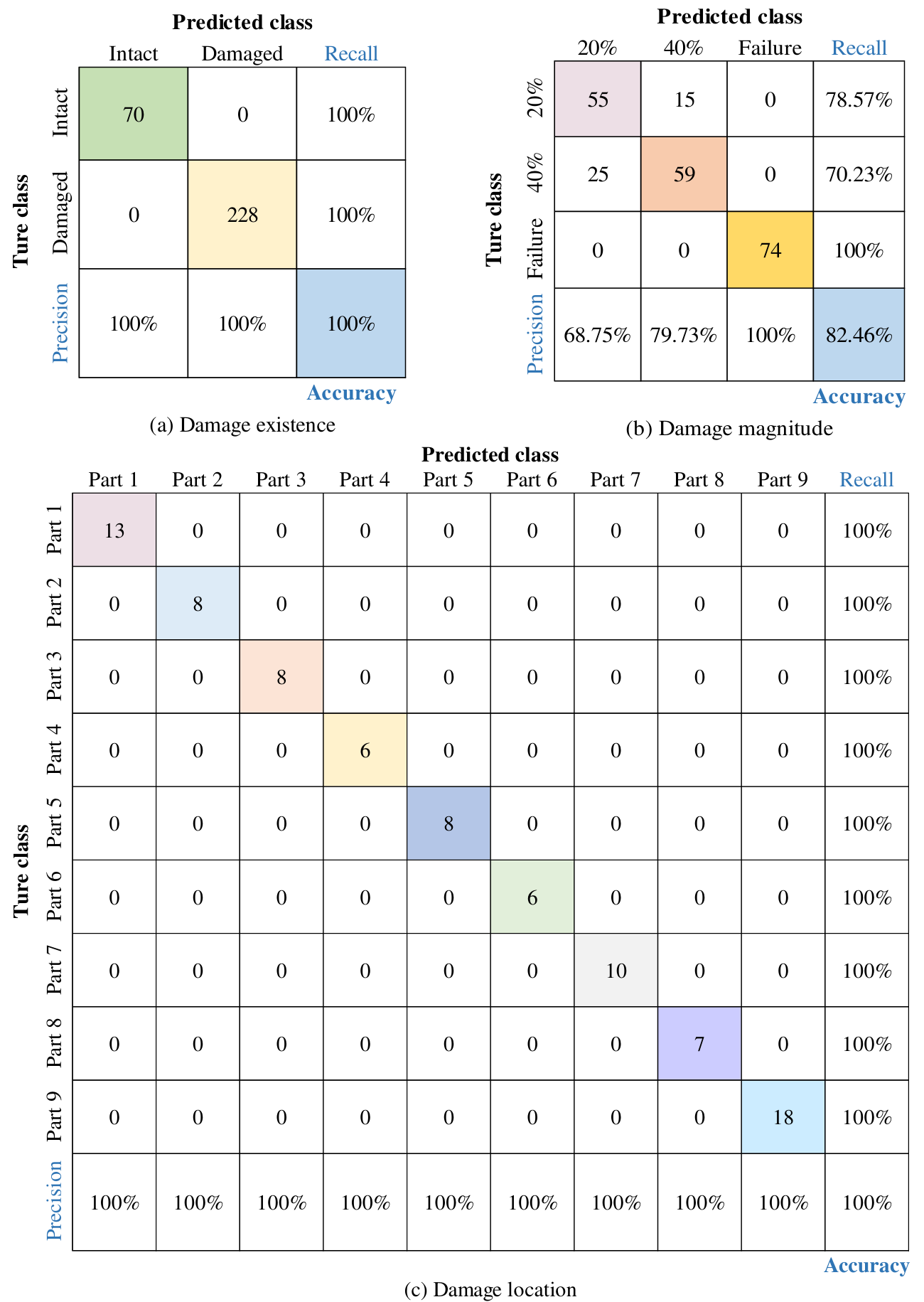}
    \caption{The damage identification results of kNN based on modal analysis under Case 2 (considering the noise level from 0\% to 7\% into FEM natural frequency): (a) Damage existence, (b) Damage magnitude, (c) Damage location}
    \label{kNN-classification-7Err}
\end{figure}

Three kNN classifiers are used sequentially to determine damage existence, damage level, and failure location, for both Case 1 and Case 2. 75\% of the samples are used to train the network, and 25\% are used for testing. The identification results of Case 1 and Case 2 are shown in Figures \ref{kNN-classification-0Err} and \ref{kNN-classification-7Err}, respectively. The k-values and distance method are determined by minimizing the cross-validation loss error, using the cross validation function "crossval" in MATLAB \cite{Matlab-documentation}, which is also a kind of hyperparameter optimization. The search space includes all the train datasets. The automatically selected k-values and distance methods for both Case 1 and Case 2 are listed in Table \ref{k-values-case1-case2}.

\begin{table}[h]\footnotesize
\centering
\caption{The automatically selected k-values and distance methods for damage identification of Case 1 and Case 2}
\label{k-values-case1-case2}
\begin{tabular}{|c|c|c|c|c|}
\hline
\text{Description} & \multicolumn{2}{|c|}{Case 1} & \multicolumn{2}{|c|}{Case 2}\\ \hline
\text{Parameter}&\text{k value} &\text{Distance} & \text{k value} &\text{Distance} \\ \hline
\text{Damage existence}&\text{7} &\text{Chebyshev} & \text{9} &\text{Chebyshev} \\
\text{Damage magnitude}&\text{4} &\text{Chebyshev} & \text{1} &\text{Minkowski}  \\
\text{Damage location}&\text{1} &\text{Chebyshev} & \text{3} &\text{Euclidean} \\
\hline
\end{tabular}
\end{table}

The confusion matrix provides a comprehensive way to understand the strengths and weaknesses of the classification model. The results indicate that the kNN algorithm achieved 100\% accuracy in determining damage existence and failure location. However, the accuracy of the damage state in Case 2 is lower than in Case 1, with confusion occurring within the 20\% and 40\% datasets. This is primarily because the deviation of some natural frequency values, caused by the error within 7\%, is greater than the deviation caused by different damage degrees. Fortunately, this does not affect the identification of failure damage, which is crucial for bridge damage detection.

\subsection{Conclusion}\label{Conclusion-kNN}

This section employs the first 14 natural frequencies of the KW51 bridge, derived from modal analysis results, as features for damage identification using kNN algorithms. The damage identification process is divided into three steps: damage existence identification, damage magnitude detection, and damage location identification. The identification results indicate that the bridge modes are highly sensitive to arch damage. The damage magnitude is precisely divided into three levels: 20\%, 40\% and Failure, with civil engineering significance. The damage degree of the bridge structure can be accurately identified by the kNN algorithm. Furthermore, the kNN algorithm is used to locate the failure damage. A comprehensive program of damage quantification and damage location based on modal analysis is verified in this section, which is of great important for the long-term monitoring of civil structures.

However, modal analysis requires extensive long-term monitoring data, reducing detection efficiency. Therefore, an immediate-use damage detection strategy based on acceleration signals needs to be proposed.

\section{Damage identification based on dynamic analysis of CMLDI method}\label{Section7-acceleration}

\begin{figure}[!htb]
    \centering
    \includegraphics[width=0.65\textwidth]{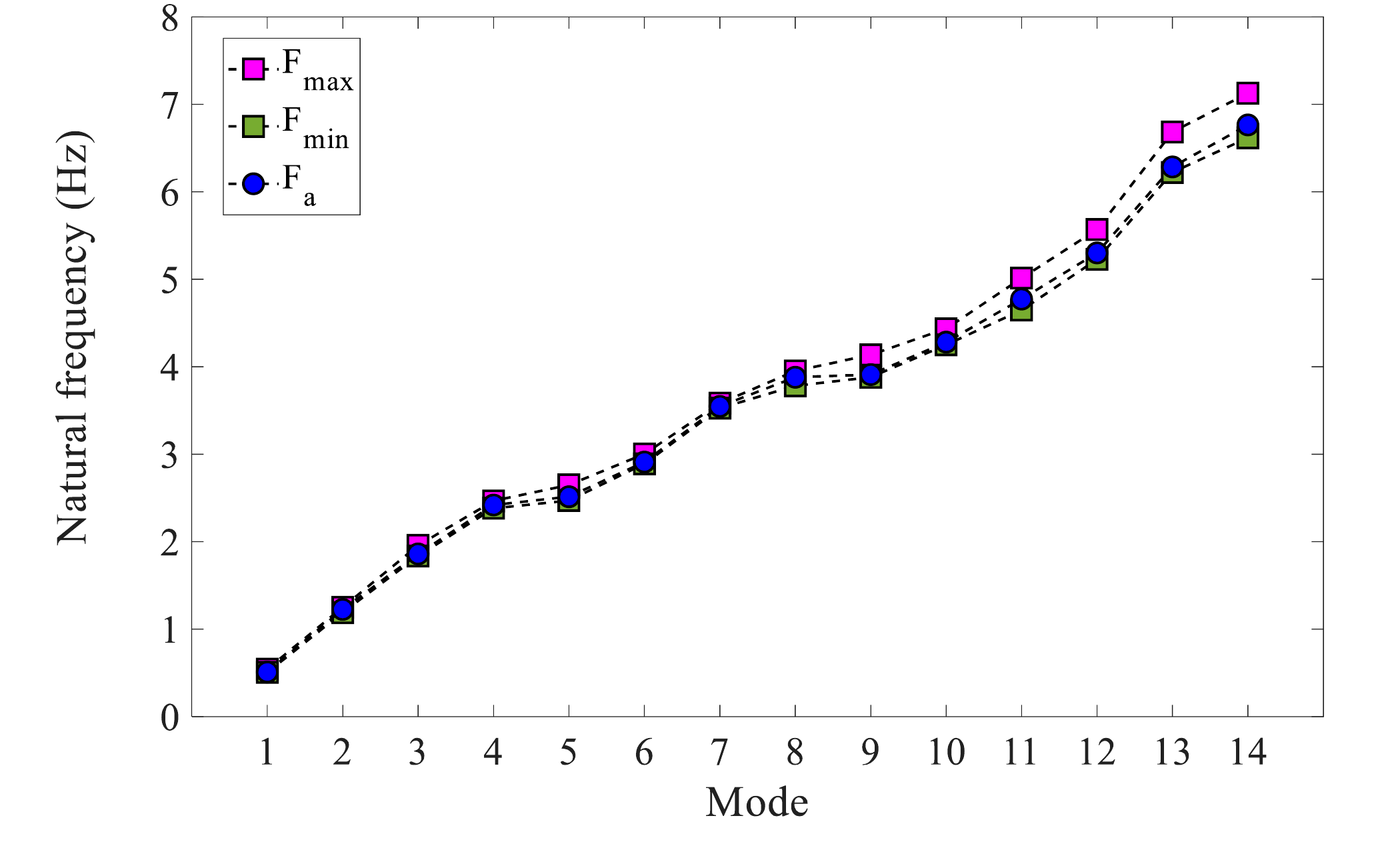}
    \caption{Comparison of train pass acceleration and the ambient noise from measured data}
    \label{MeasuredAcc-Ambient}
\end{figure}

Different from the method based on modal analysis input, the input of forced acceleration signals come from the train-passing period. Three accelerations are obtained for the damage identification process of the KW51 bridge, the locations of the force acceleration are defined according to the sensor positions on the arch of KW51 bridge. This section focuses on lateral accelerations. Each acceleration is within a 13-second interval, assuming the speed range of the train is 15m/s to 30m/s. Datasets collected from monitoring between October 2019 and December 2019 are utilized as "Intact" class. The amplitude of the ambient noise in the measured data is less than 1\% of the acceleration amplitude, as shown in Figure \ref{MeasuredAcc-Ambient}. 

\begin{figure}[!htb]
    \centering
    \includegraphics[width=0.65\textwidth]{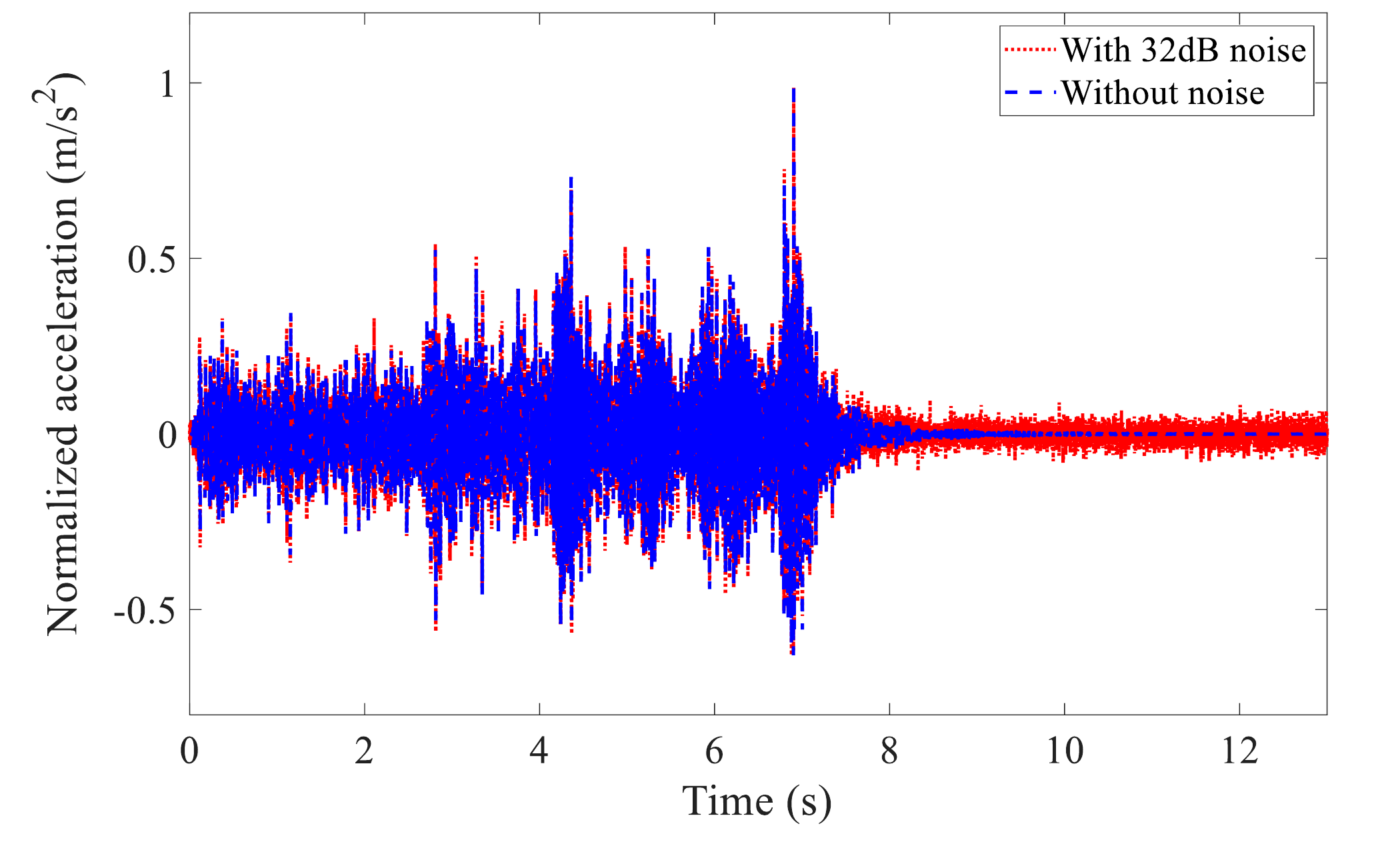}
    \caption{Comparison of FEM accelerations with 32dB noise and without noise}
    \label{Acc-noise-32dB}
\end{figure}

FEM also obtains 6 samples of acceleration response for the intact state at different train speeds. Accelerations for damaged structure are obtained through FEM and categorized into three damage levels: 20\%, 40\%, and Failure. Damage levels of 5\%, 10\%, 15\% and 20\% are labeled as 20\%, while damage levels of 25\%, 30\%, 35\%, 40\% are classified as 40\%. Failure label comes from FEM by removing the element from the failure part. The Gaussian noise with 32dB signal-to-noise ratio (SNR) is considered in acceleration responses. The definition of SNR is:
\begin{equation}\label{SNR}
\begin{aligned}
    \boldsymbol{SNR} = 10\log_{10}{\frac{\Vert \boldsymbol{A} \Vert_2}{\Vert \boldsymbol{G}_n \Vert_2}}
\end{aligned}
\end{equation}

\noindent where $ \Vert \Vert_2 $ is Euclidian 2-norm, $\boldsymbol{A} $ is the original acceleration, and $\boldsymbol{G}_n $ is the Gaussian noise. \\
Six speeds of moving load are considered to simulate the train velocities: 15.75m/s, 18m/s, 20m/s, 21m/s, 25m/s and 27m/s. Figure \ref{Acc-noise-32dB} compares acceleration curves with 32dB noise and without noise. The 32dB noise level is much higher than the 1\% ambient noise level from the measured data. To accomplish various stages of damage identification, the machine learning methods of stacked GRU, kNN, and CNN methods are employed.

The dataset comprises 1424 samples, with 236 labeled as 'Intact' and 1188 labeled as 'Damaged', including 432 labeled as 'Failure'. Stacked GRU and kNN algorithms are utilized for identifying the existence and magnitude of damage. Stacked GRU is employed to distinguish between intact and damaged structures, while kNN is utilized to classify the degree of damage. Additionally, CNN is deployed to classify images obtained from wavelet transform in failure cases, facilitating the identification of the failure location. Table \ref{Total samples accelerations} shows the description of the accelerations samples.

\begin{table}[h]\footnotesize
\centering
\caption{The description of the samples of accelerations}
\label{Total samples accelerations}
\begin{tabular}{|c|c|c|c|c|}
\hline
 \text{Noise level} &\text{Description} & \text{Label}&\text{Source}&\text{Sample}\\ \hline
\text{ 1\%, 32dB}&\text{Intact}&\text{Intact}&\text{Measured, FEM}&\text{236}\\
\text{32dB}&\text{Damage levels: 5\%, 10\%, 15\%, 20\%}&\text{20\%}&\text{FEM}&\text{378}\\
\text{32dB}&\text{Damage levels: 25\%, 30\%, 35\%, 40\%}&\text{40\%}&\text{FEM}&\text{378}\\
\text{32dB}&\text{Failure}&\text{Failure}&\text{FEM}&\text{432}\\
\hline
\end{tabular}
\end{table}

\subsection{Stacked GRU and kNN algorithm for damage magnitude}\label{Acceleration damage magnitude}

To validate the applicability of ML models, six different cases are considered, as described in Table \ref{Test cases}. The performance of the model is evaluated using the test set, and the results are presented in the form of a confusion matrix. 
\begin{table}[h]\footnotesize
\centering
\caption{Description of test cases}
\label{Test cases}
\begin{tabular}{|c|c|}
\hline
\text{Train speeds (m/s)} & \text{Test speed (m/s)}\\ \hline
\text{18, 20, 21, 25, 27} &\text{15.75} \\
\text{15.75, 20, 21, 25, 27} &\text{18} \\
\text{15.75, 18, 21, 25, 27} &\text{20} \\
\text{15.75, 18, 20, 25, 27} &\text{21} \\
\text{15.75, 18, 20, 21, 27} &\text{25}  \\
\text{15.75, 18, 20, 21, 25} &\text{27} \\
\hline
\end{tabular}
\end{table}

Considering the same dataset, the comparison of training time and memory costing for stacked GRU (stack number is 200) and kNN is shown in Table \ref{Compare-memory}. The training time of stacked GRU is slightly shorter than kNN, and stacked GRU network requires less memory. The stacked GRU network file is only 704KB, while the kNN network requires 40,564KB. This difference becomes more significant as the dataset grows. Thus, the stacked GRU is employed to identify the existence of structural damage rather than kNN algorithm.
\begin{table}[h]\footnotesize
\centering
\caption{The comparison of training time and memory costing of stacked GRU (stack number is 200) and kNN for damage existence identification}
\label{Compare-memory}
\begin{tabular}{|c|c|c|}
\hline
\text{Description} & \text{stacked GRU} & \text{kNN}\\ \hline
\text{Memory} & \text{704 kB} & \text{40564 kB}\\
\text{Training time} & \text{42 s} & \text{56 s}\\
\hline
\end{tabular}
\end{table}
 
\begin{figure}[!htb]
    \centering
    \includegraphics[width=0.89\textwidth]{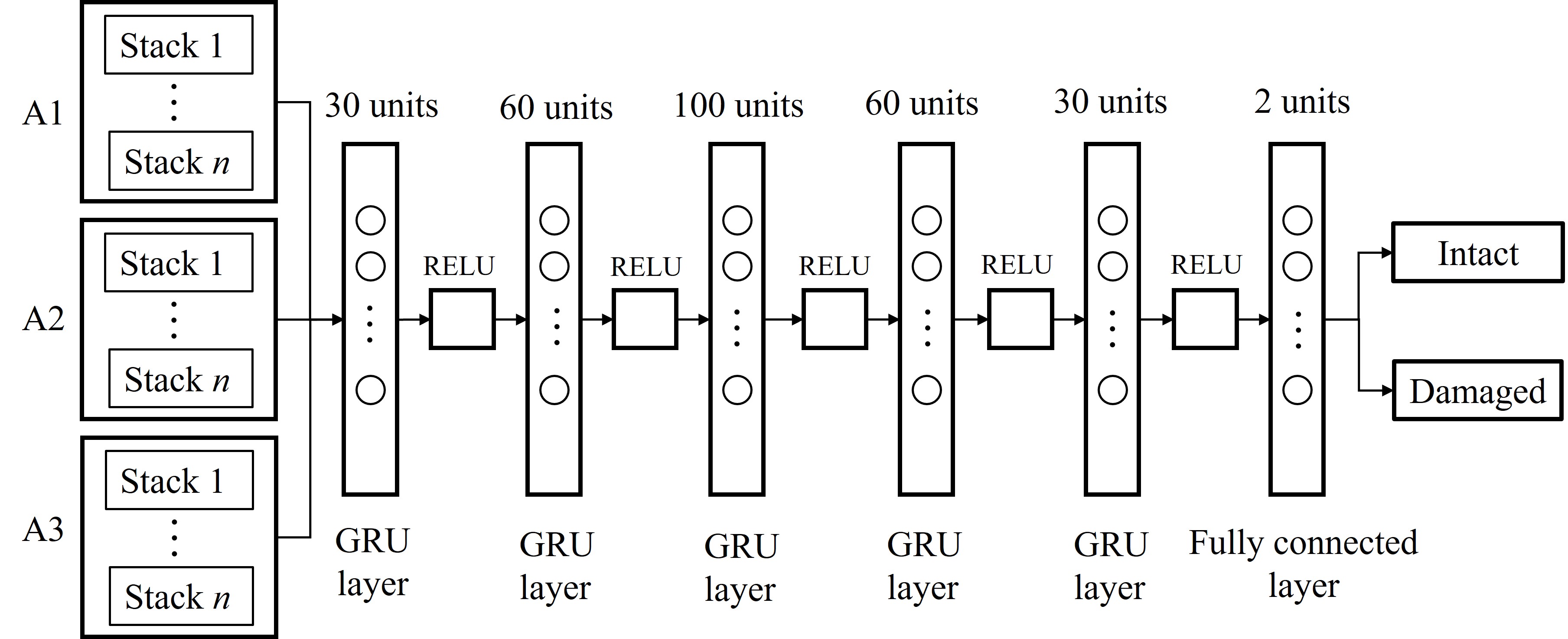}
    \caption{Multilayer stacked GRU architecture with hidden units }
    \label{StackedGRUnetworks}
\end{figure}

\begin{figure}[!htb]
    \centering
    \includegraphics[width=0.62\textwidth]{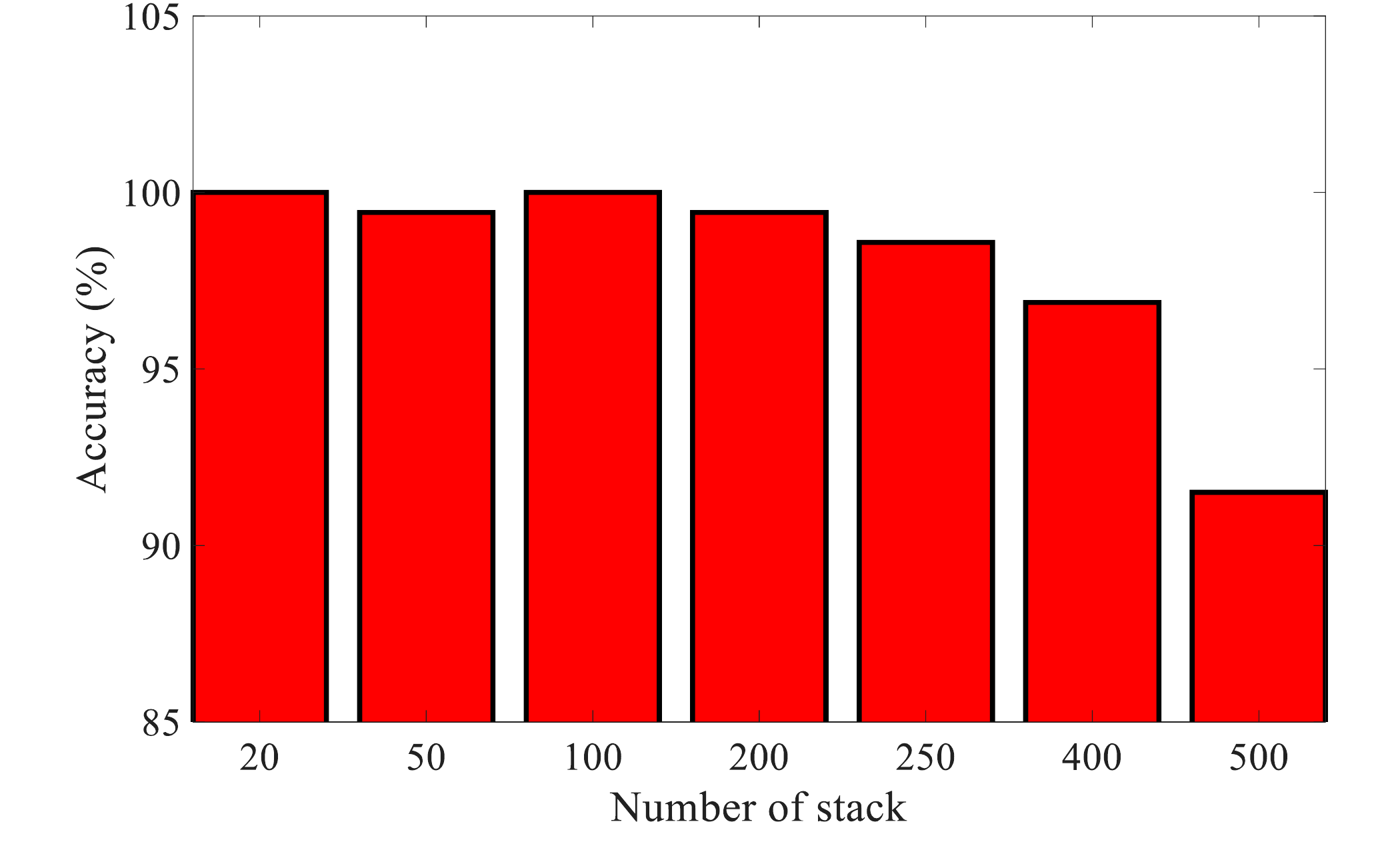}
    \caption{The accuracy of damage existence identification of stacked GRU with multiple numbers of stack}
    \label{Accuracy-GRU-class2}
\end{figure}

\begin{figure}[!htb]
    \centering
    \includegraphics[width=0.62\textwidth]{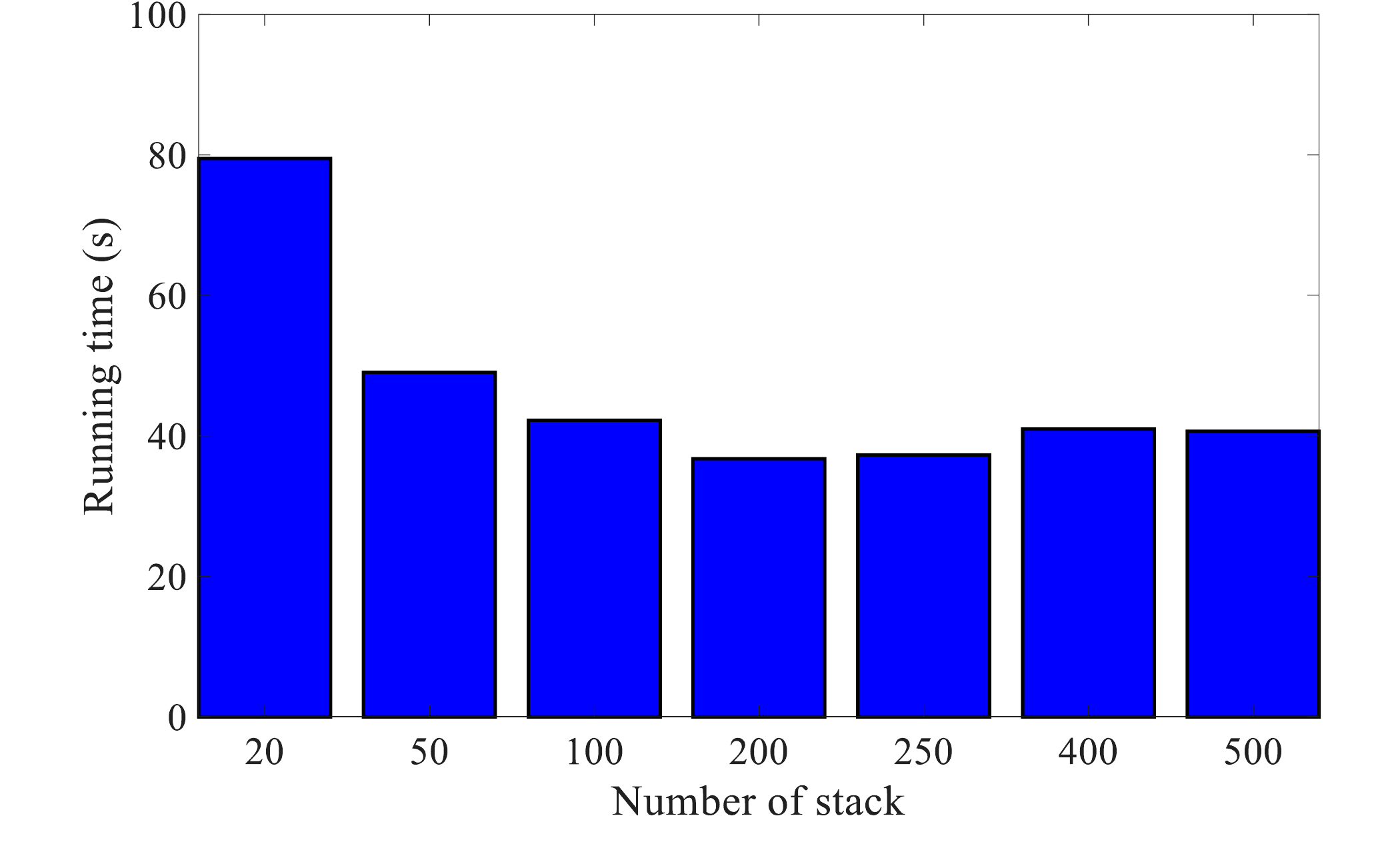}
    \caption{The running time of damage existence identification of stacked GRU with multiple numbers of stack}
    \label{Time-GRU-class2}
\end{figure}

The network of stacked GRU is shown in Figure \ref{StackedGRUnetworks}. An acceleration time series with 10,000 time steps is divided into multiple stack numbers: 20, 50, 100, 200, 250, 400, and 500. The number of stacks to accuracy and training time curves are presented in Figure \ref{Accuracy-GRU-class2} and \ref{Time-GRU-class2} using a total of 1424 samples, with 25\% samples to predict. It can be seen that both the accuracy of the stacked GRU classification and the running time for training the network decrease with the increasing number of stacks. To balance the running time and accuracy, the number of stacks is defined as 200 for classifying the accelerations of the KW51 bridge.

\begin{figure}[!htb]
    \centering
    \includegraphics[width=1\textwidth]{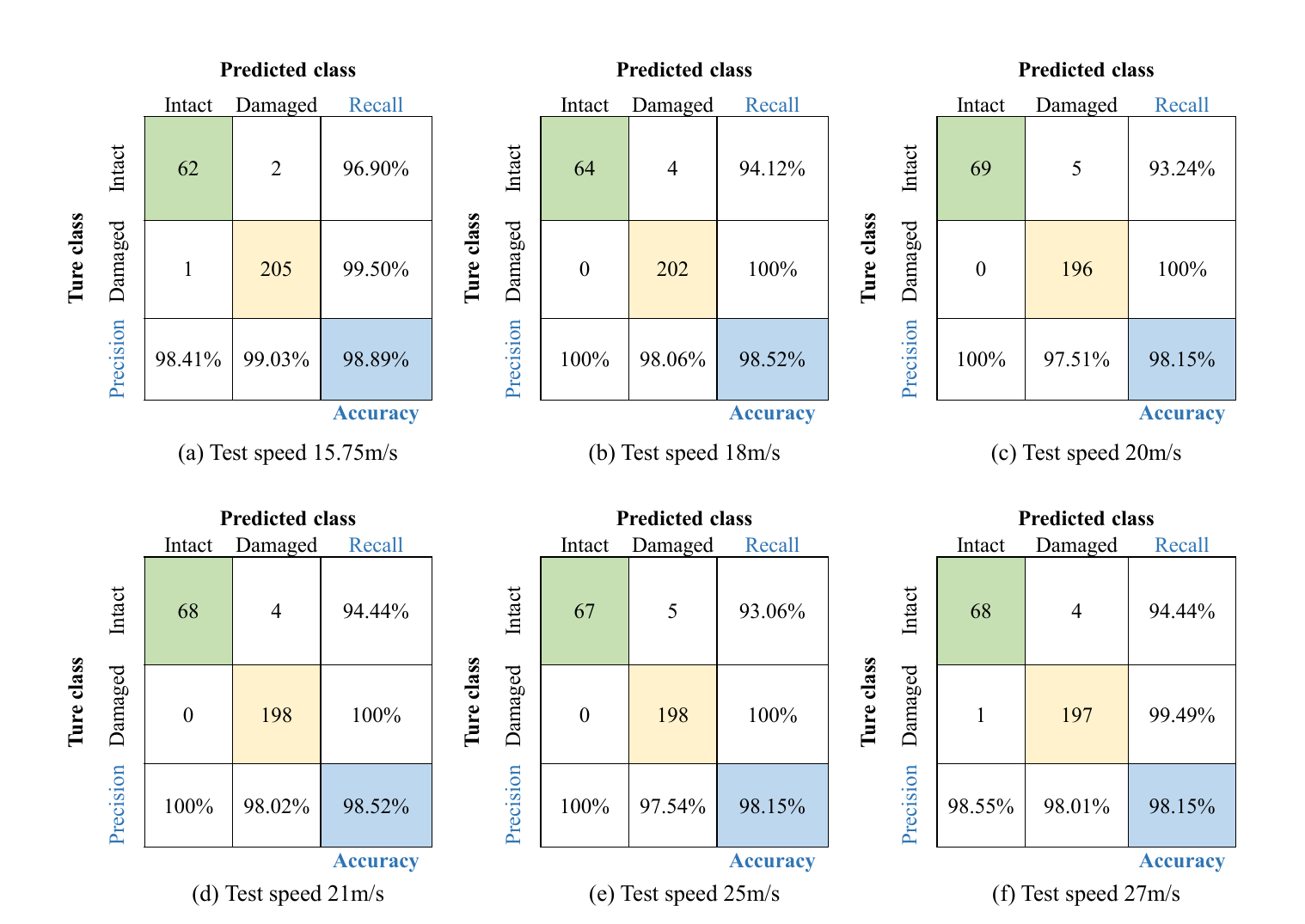}
    \caption{The damage existence identification results of stacked GRU with stack number of 200, based on dynamic analysis in six test cases: (a) Test speed 15.75m/s, (b) Test speed 18m/s, (c) Test speed 20m/s, (d) Test speed 21m/s, (e) Test speed 25m/s, (f) Test speed 27m/s}
    \label{GRU-speeds}
\end{figure}

To validate the applicability of stacked GRU, the above six test cases are considered. The performance of the model is evaluated using the test set, and the results are presented in the form of a confusion matrix, as shown in Figure \ref{GRU-speeds}. The confusion matrix displays two classes (Intact, and Damaged). The performance of stacked GRU can be assessed using the third row for precision and the third column for recall. Based on the results shown in Figure \ref{GRU-speeds}, all test cases achieved an accuracy rate exceeding 98\%, demonstrating the quality of the stacked GRU for classifying damage existence. Additionally, the results indicate that the majority of mis-classifications occur with the intact sample data, suggesting that the network accurately classifies the damaged samples. This is particularly beneficial for the bridge damage detection system.

\begin{figure}[!htb]
    \centering
    \includegraphics[width=0.62\textwidth]{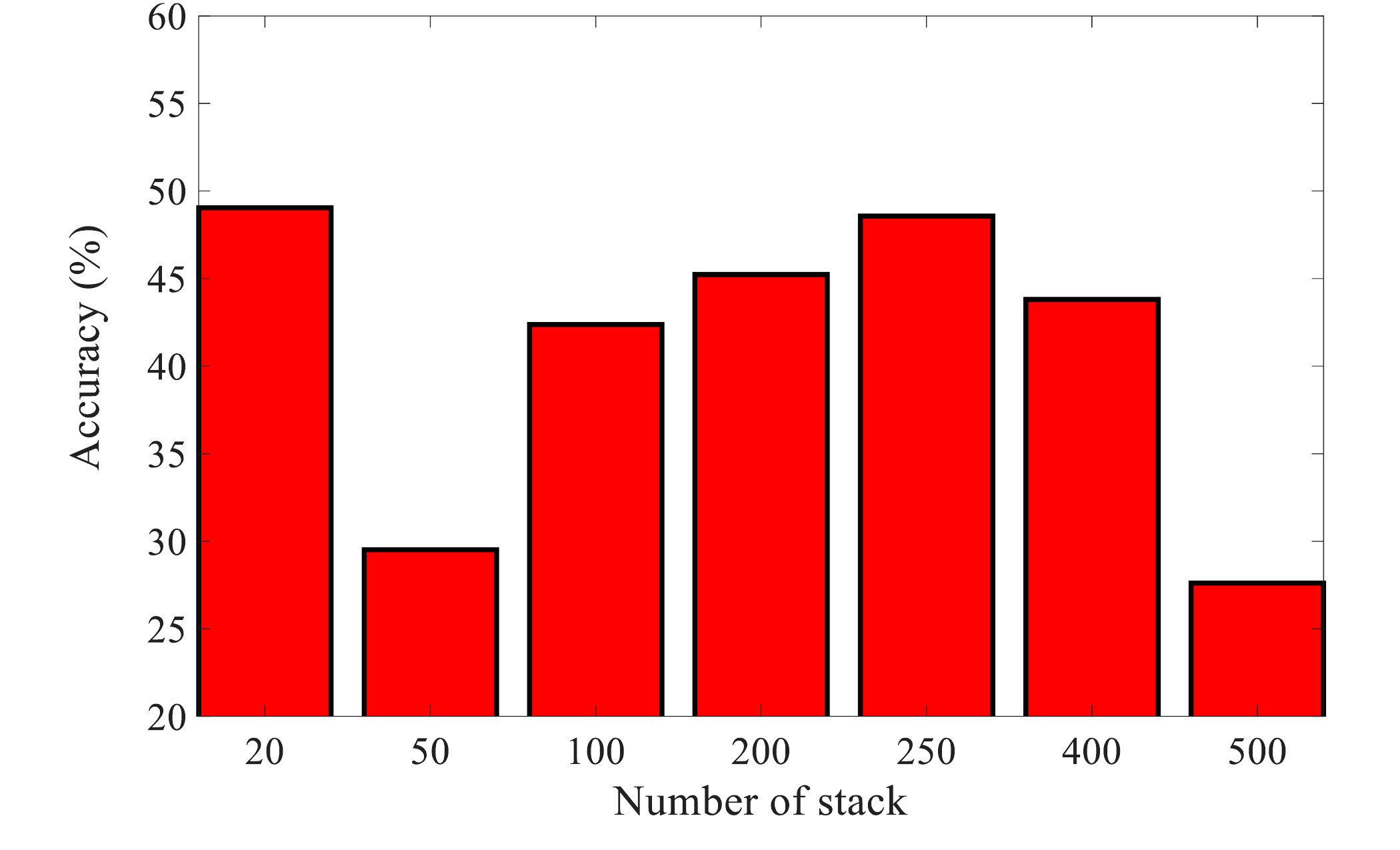}
    \caption{The accuracy of damage magnitude identification of stacked GRU using different stack numbers with the test speed of 15.75m/s}
    \label{GRU_extent_accuracy_out15ms}
\end{figure}

\begin{figure}[!htb]
    \centering
    \includegraphics[width=1\textwidth]{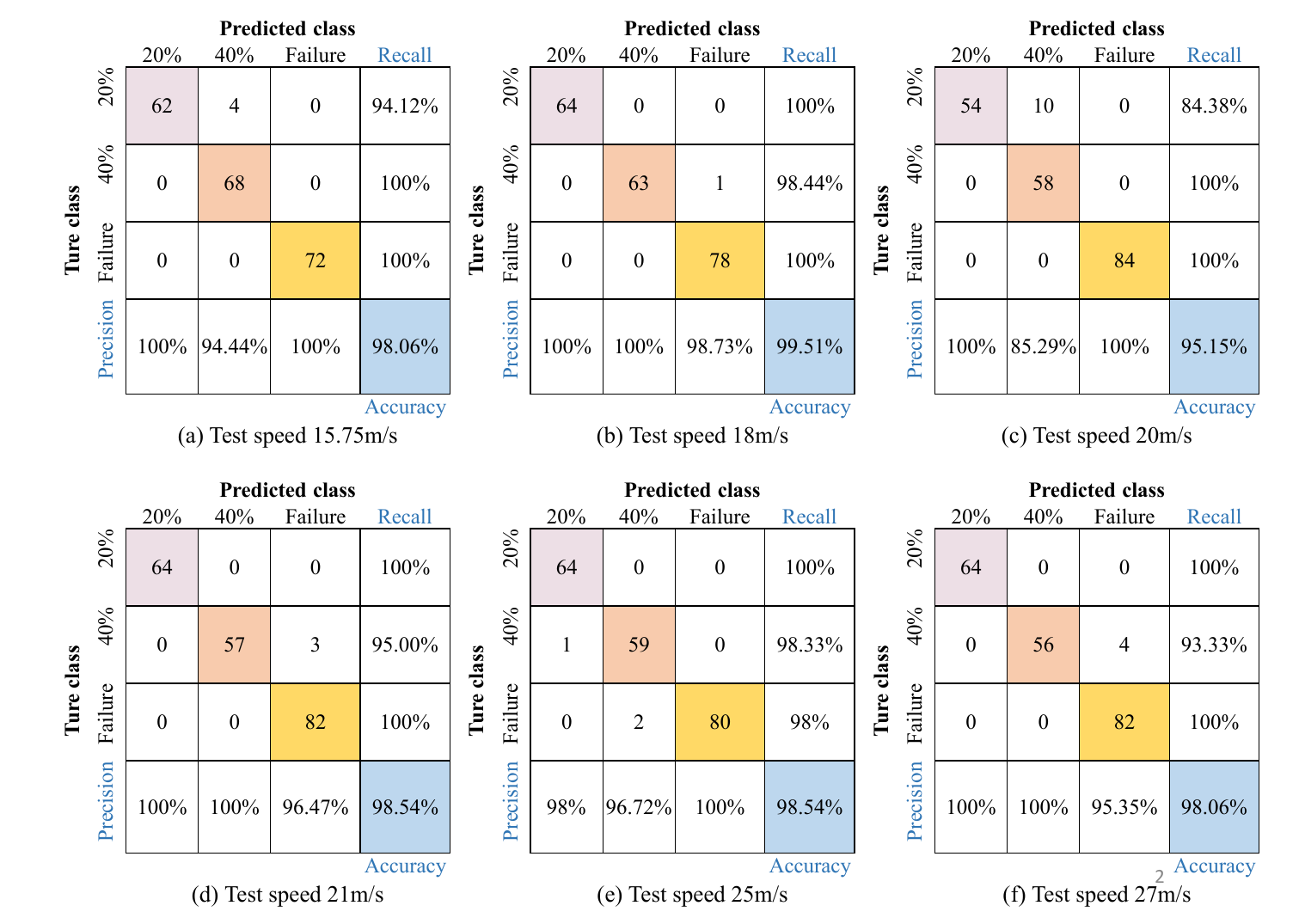}
    \caption{The damage magnitude identification results of kNN classifier based on dynamic analysis in six test cases: (a) Test speed 15.75m/s, (b) Test speed 18m/s, (c) Test speed 20m/s, (d) Test speed 21m/s, (e) Test speed 25m/s, (f) Test speed 27m/s}
    \label{KNN-speeds}
\end{figure}

Although stacked GRU performs well in determining the existence of damage, its ability to identify the magnitude of damage is not satisfying. The accuracy of stacked GRU using different stack numbers with the test speed of 15.75m/s is shown in Figure \ref{GRU_extent_accuracy_out15ms}. All the accuracy values are lower than 50\%. 
The kNN algorithm shows an advantage in classifying the degree of damage. The damage-state data filtered by the stacked GRU is fed as input for the kNN algorithm. The input dataset for kNN is preprocessed by the stacked GRU, significantly reducing the memory requirements for kNN. Therefore, kNN algorithm is arranged to identify the damage magnitude with classes of 20\%, 40\%, and failure. The kNN algorithm utilizes the frequency features that can be obtained by Fourier transform. The frequency band of 10Hz to 150Hz is considered. The above six test cases are considered for kNN damage identification. The damage magnitude results identified by the kNN algorithm are shown in the Figure \ref{KNN-speeds}. All accuracy of six test cases are higher than 98\% except for the test speed of 20m/s with an accuracy of 95.15\%. All the confusion (lower accuracy) data comes from datasets with adjacent labels, especially between 20\% and 40\%. The 20\% label includes damage levels from 5\% to 20\%, and the 40\% label covers damage levels from 25\% to 40\%. Because the natural frequencies of the 20\% and 25\% damage levels are similar, the kNN classifier often experiences confusion between these two cases. The 20\% label includes damage levels from 5\% to 20\%, and the 40\% label covers damage levels from 25\% to 40\%. Because the natural frequencies of the 20\% and 25\% damage levels are similar, the kNN classifier often experiences confusion between these two cases. However, it can achieve 100\% accuracy for the identification results of failure samples. The values of k for the kNN are listed in the Table \ref{kNN-k-values}.

\begin{table}[h]\footnotesize
\centering
\caption{The automatically selected k-values and distance methods for damage magnitude identification using kNN}
\label{kNN-k-values}
\begin{tabular}{|c|c|c|}
\hline
\text{Test speed} & \text{k value} & \text{Distance}\\ \hline
\text{15.75m/s} & \text{1} & \text{Chebyshev}\\
\text{18m/s} & \text{1} & \text{Minkowski}\\
\text{20m/s} & \text{1} & \text{Chebyshev}\\
\text{21m/s} & \text{1} & \text{Euclidean}\\
\text{25m/s} & \text{1} & \text{Minkowski} \\
\text{27m/s} & \text{1} & \text{Euclidean}\\
\hline
\end{tabular}
\end{table}

The advantages of stacked GRU and kNN algorithms in the problems of damage existence and damage magnitude identification are respectively discussed. Both ML methods demonstrated good performance in identifying the presence and severity of damage with six test cases. However, these methods are limited in identifying damage location, as time series and frequency series provide limited information of damage location. Therefore, an additional approach is needed for accurate damage location identification.

\subsection{CNN method for failure location}\label{Acceleration failure location}

Once the severity of damage is identified, location identification of the failure becomes crucial. In this paper, displacement-frequency images of 432 failure samples are obtained through wavelet transform, as described in Section 4 (Figure \ref{wavelet-A3}). The Morlet wavelet basis function is used in MATLAB for this transformation. The acceleration signal features at multiple moving speeds are normalized by converting the horizontal axis of the images to distance.

\begin{figure}[!htb]
    \centering
    \includegraphics[width=0.8\textwidth]{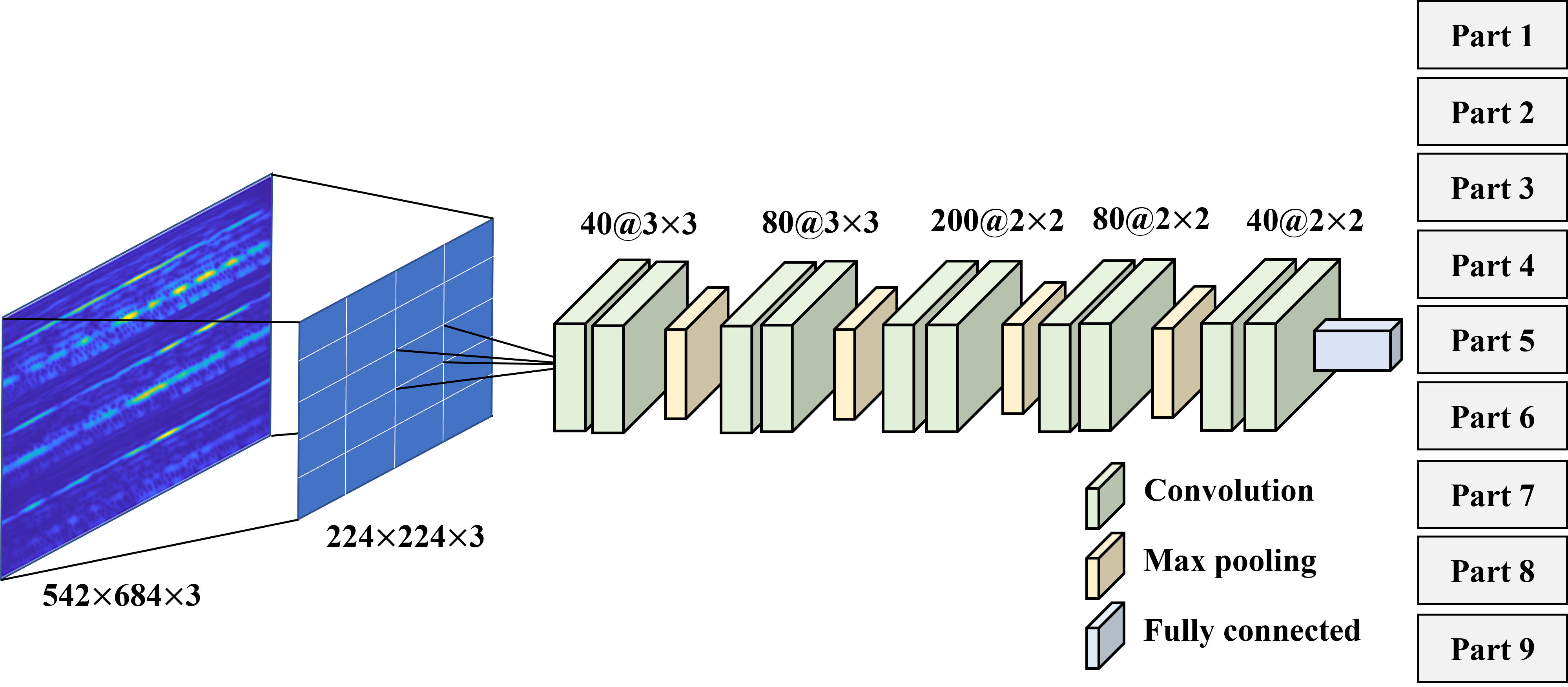}
    \caption{CNN architecture for image classification}
    \label{CNN network}
\end{figure}

A simple CNN is used for damage location identification, following the finding of W. Liao et al. \cite{simpleCNN}, who demonstrated that a simple CNN outperforms more complex architectures when using wavelet transform images. The CNN employed in this paper is shown in Figure \ref{CNN network}. The input of the CNN consists of resized images with dimensions of $224\times224\times3$, and the nine output classes represent the damage locations on the arch of the KW51 bridge.

\begin{figure}[!htb]
    \centering
    \includegraphics[width=0.62\textwidth]{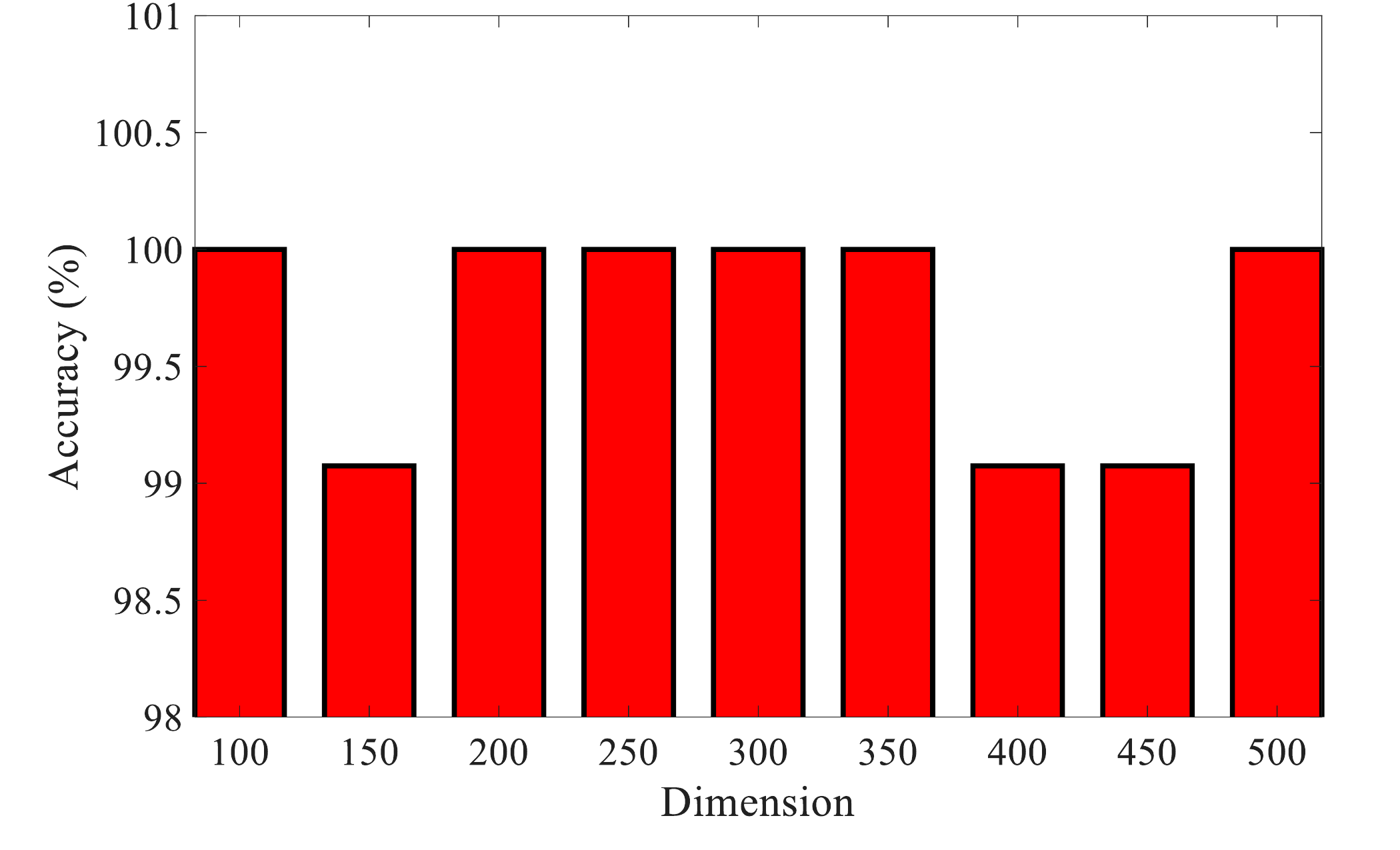}
    \caption{The accuracy of CNN for damage location identification with different resized image dimensions}
    \label{Accuracy-CNN-location9}
\end{figure}

\begin{figure}[!htb]
    \centering
    \includegraphics[width=0.62\textwidth]{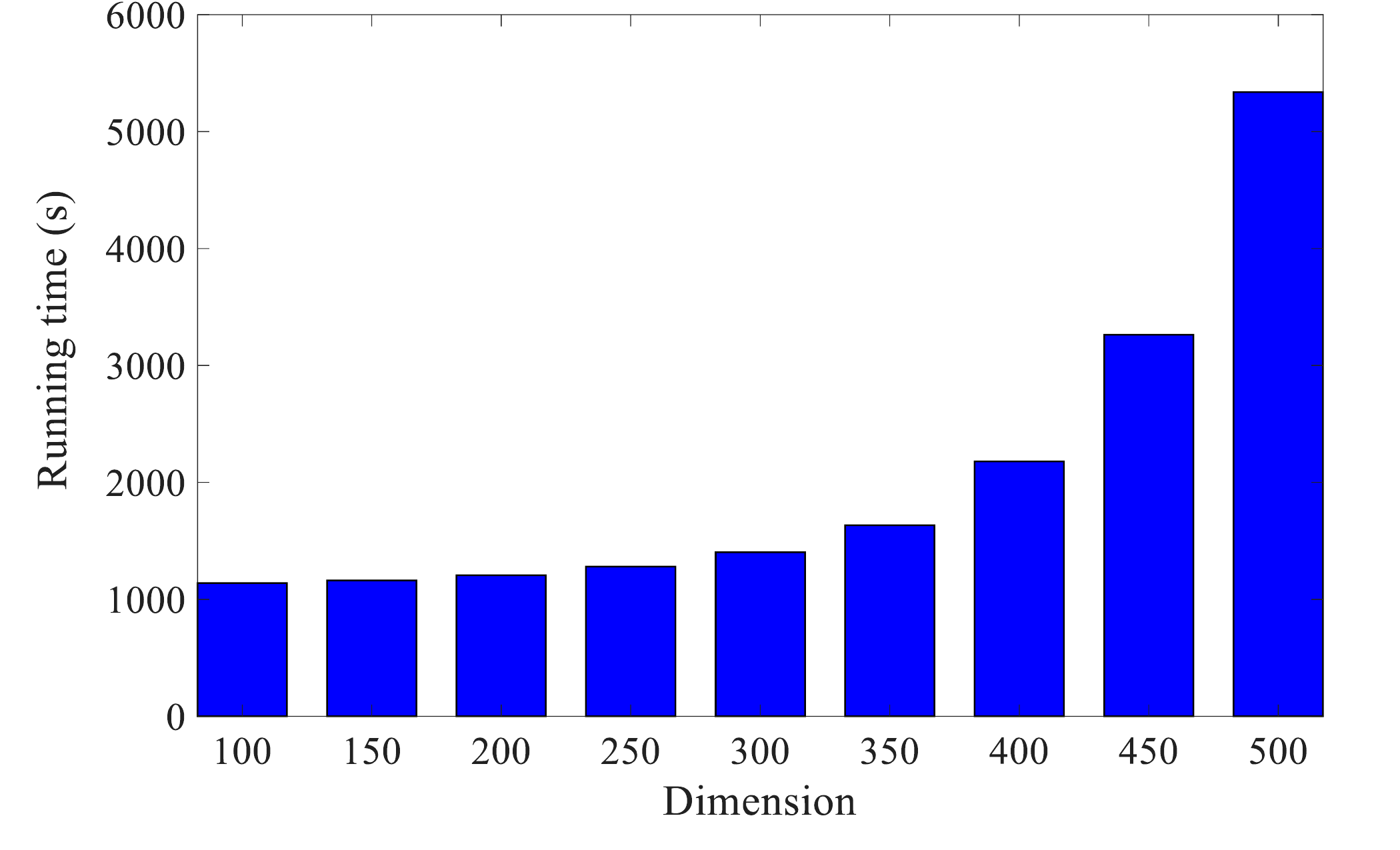}
    \caption{The training time of CNN for damage location identification with different resized image dimensions}
    \label{Time-CNN-location9}
\end{figure}

The dimension of the resized image is defined based on Figure \ref{Accuracy-CNN-location9} and \ref{Time-CNN-location9}, which represents the accuracy and training time of the CNN using different resize dimensions ranging from $100\times100\times3$ to $500\times500\times3$. The results show that the identification accuracy is high (over 99\%), but the running time significantly increases with the rise in resize dimensions. Thus, $224\times224\times3$ is the suitable size for the CNN to identify the damage location.

\begin{figure}[!htb]
    \centering
 \includegraphics[width=1.0\textwidth]{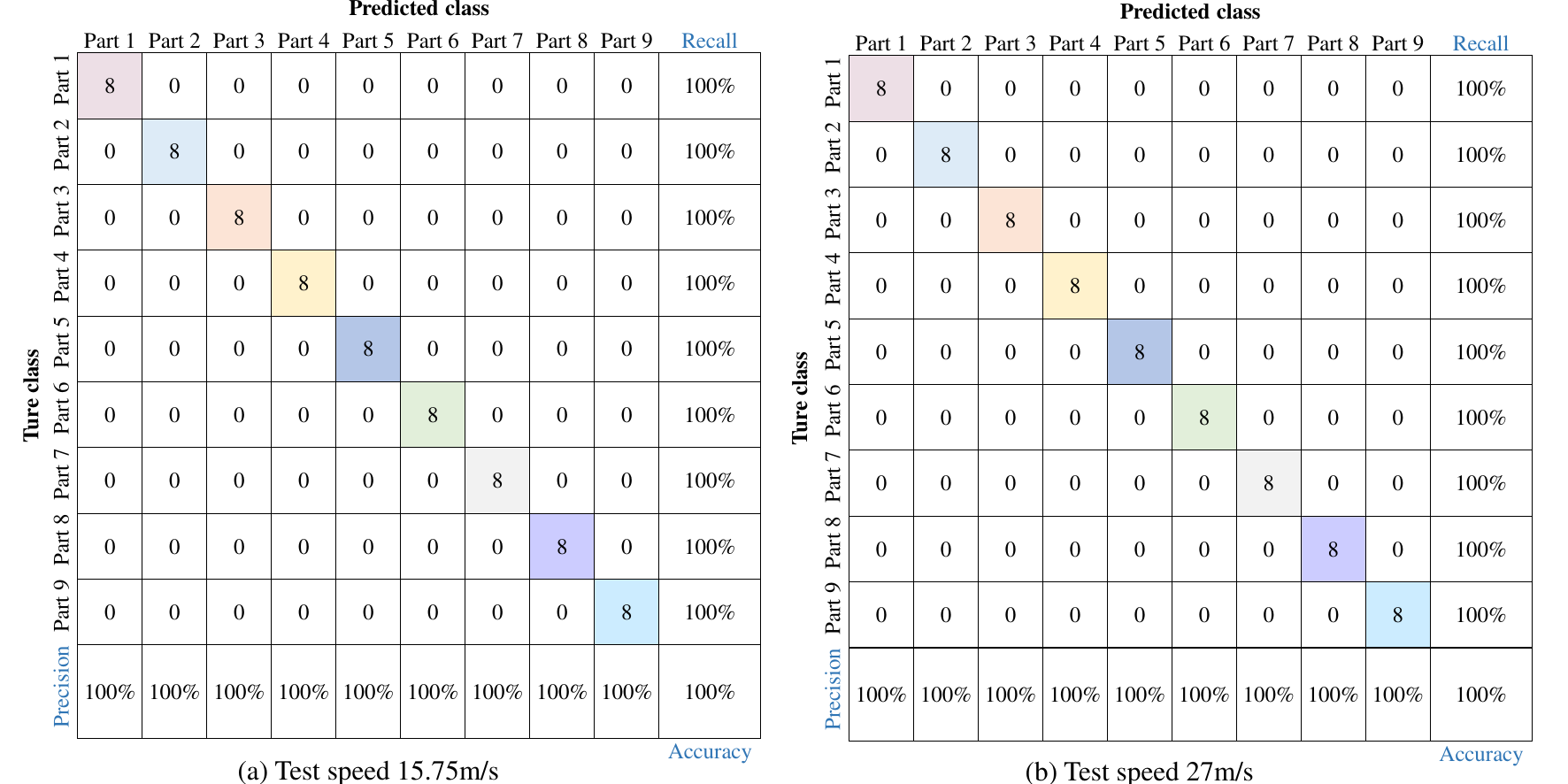}
    \caption{The damage location identification results of CNN with test speeds of 15.75m/s and 27m/s based on dynamic analysis: (a) Test speed 15.75m/s, (b) Test speed 27m/s}
    \label{CNN-location-speeds-acc}
\end{figure}

The CNN method is used to classify the six test cases to identify the failure location, achieving an accuracy of 100\%. Figure \ref{CNN-location-speeds-acc} shows the classification results for the test speeds of 15.75 m/s and 27 m/s, demonstrating the effectiveness of the CNN method in failure location identification.

\subsection{Conclusion}\label{Conclusion-Acceleration}

This section utilizes the damage identification method based on acceleration responses. The accelerations are processed by different techniques, time series stacking, Fourier transform, and wavelet transform, to obtain different acceleration features. The 32dB noise level is considered to simulate real-world conditions. A one-to-one machine learning approach is adopted for these damage-sensitive features. Stacked GRU is used to process the stacked time series data, significantly improving network training time and achieving over 98\% accuracy in classifying damage existence. Unlike modal analysis input, the frequency features obtained through the Fourier transform form a sequence within the 10Hz-150Hz range. The kNN algorithm is employed to classify these frequency sequences for damage degree identification, with an accuracy over 95\%. However, both stacked time series and frequency features are less effective in identifying damage locations. To address this, time-frequency features of the acceleration are used in the form of images by wavelet transform. The horizontal axis is normalized to distance to account for varying train speeds. A CNN method is adopted to classify the displacement-frequency images, achieving a damage location accuracy of 100\%.

In conclusion, the proposed damage identification method based on acceleration presents robustness and efficiency by combining multiple ML approaches rather than relying on a single ML method. The results demonstrate the stability and effectiveness of the proposed method in accurately detecting and locating bridge damage. In the future, this damage identification method can be easily integrated into a user-friendly interface for monitoring and maintenance planning for result visualization.

\section{Discussion, conclusions, and future directions}\label{Section8-Discussion}

This paper proposes the CMLDI method, which aims to address the need for damage detection in civil structures under various conditions rather than sporadic evaluations of single damage identification scenarios. The CMLDI method offers a solution for comprehensive damage identification in practical full-scale bridge structures by combining customized signal processing methods with ML algorithms. Damage identification based on modal analysis is employed for long-term monitoring. When short-term error identification is necessary, the CMLDI method utilizes a damage identification strategy based on time history data, with several seconds of sampling time. Environmental noise is taken into account in both the modal analysis features and the acceleration responses; the noise level is determined by the measured data. The current approach assumes a fixed loading case.

The damage identification strategy of the CMLDI method based on modal analysis input employed three kNN classifiers sequentially to determine the existence, magnitude, and location of the damage. The results show that the kNN classifier achieves 98\% accuracy in identifying the presence of damage and the failure location when a 7\% uncertainty level is included. The confusions mainly occur within the moderate damage levels, especially due to the definition of 20\% type and 40\% type. However, the identification results were mostly unaffected, underscoring the reliability and robustness of the proposed system for practical bridge damage detection applications.

The damage identification strategy of CMLDI method also utilizes short-term acceleration time histories. For different damage detection requirements, the feature extraction methodology through signal processing is introduced, and the pertinent machine learning approach is presented in detail. Three different signal processing techniques are considered: time series stacking, Fourier transform, and Wavelet transform. Even with the incorporation of noise, the combined approach proves to be effective. The time-stacked features combined with the stacked GRU algorithm demonstrate strong performance in identifying the existence of damage. Compared to the kNN algorithm, it requires less training time and storage memory. However, the information on the magnitude of damage contained in the time-stacked sequences is insufficient, making it challenging to accurately identify the severity of the damage. In contrast, kNN shows higher accuracy in determining the damage magnitude by processing frequency sequences. Both stacked GRU and kNN show limited effectiveness in identifying the location of damage. Wavelet transform is used to generate time-frequency images of the acceleration signals since the time series and frequency sequence are not particularly sensitive to damage location. By converting the horizontal axis of these images to distance, the acceleration signals are normalized for different train speeds. A CNN method is then applied to classify the distance-frequency images, achieving 100\% accuracy damage location identification results.

In conclusion, this paper presents an effective CMLDI method by integrating both modal analysis and dynamic analysis strategies with multiple ML methods, considering different damage sensitivity characteristics instead of single feature to detect the damage. It comprehensively considers the damage existence, magnitude, and location. The results demonstrate the stability, reliability, and high accuracy of the proposed methods in detecting and locating bridge damage. This dual-strategy approach ensures robust monitoring and precise identification of structural damage, making it a valuable tool for maintaining the integrity and safety of bridge structures. The presented approach can be directly applied and extended to various mechanical and civil engineering applications.

\section*{Acknowledgment}
This work was partially supported by the Sand Hazards and Opportunities for Resilience, Energy, and Sustainability (SHORES) Center, funded by Tamkeen under the NYUAD Research Institute Award CG013, and supported by Sandooq Al Watan Applied Research and Development (SWARD), funded by Grant No.: SWARD-F22-018. This paper also supported by China Scholarship Council (CSC), funded by Grant ID: 202306830082. The authors wish to express their gratitude to the NYUAD Center for Research Computing for their provision of resources, services, and skilled personnel. The authors would like to thank Dr. Kristof Maes (KU Leuven) for his input and advice on modeling the KW51 bridge.

\section*{Credit authorship statement}
\textbf{Yuqing Qiu:} Conceptualization, Methodology, Software, Formal Analysis, Writing - Original draft, Data generation, Visualization, Supervision. \textbf{Bilal Ahmed:} Software, Formal Analysis, Writing - Review and Editing. \textbf{Diab Abueidda:} Software, Formal Analysis, Writing - Review and Editing. \textbf{Waleed El-Sekelly:} Supervision, Project administration. \textbf{Borja García de Soto:} Supervision, Project administration, Funding acquisition.  \textbf{Tarek Abdoun:} Supervision, Project administration, Funding acquisition. \textbf{Hongli Ji:} Writing - Review and Editing. \textbf{Jinhao Qiu:} Writing - Review and Editing. \textbf{Mostafa Mobasher:} Conceptualization, Methodology, Writing - Review and Editing, Supervision, Project administration, Funding acquisition.

\bibliographystyle{elsarticle-num} 
\bibliography{mybibfile}

\end{document}